\documentclass[12pt,preprint]{aastex}
\usepackage{graphicx}

\shorttitle{The cataclysmic variable period minimum}
\shortauthors{B. Willems et al.}

\begin{document}

\title{Angular momentum losses and the orbital period distribution of
  cataclysmic variables below the period gap: effects of circumbinary
  disks} 

\author{Bart Willems}
\affil{Northwestern University, Department of Physics and Astronomy, 
  2145 Sheridan Road, Evanston, IL 60208, USA}
\email{b-willems@northwestern.edu}

\author{Ulrich Kolb} 
\affil{Department of Physics and Astronomy, The Open University,
  Walton Hall, Milton Keynes, MK7 6AA, UK}
\email{u.c.kolb@open.ac.uk}

\author{Eric L. Sandquist} 
\affil{Department of Astronomy, San Diego State University, 5500
  Campanile Drive, San Diego, CA 92182, USA}
\email{erics@mintaka.sdsu.edu}

\author{Ronald E. Taam}
\affil{Northwestern University, Department of Physics and Astronomy,
  2131 Tech Drive, Evanston, IL 60208, USA}
\email{r-taam@northwestern.edu}

\and

\author{Guillaume Dubus}
\affil{Laboratoire Leprince-Ringuet, Ecole Polytechnique, 91128
  Palaiseau, France \and Institut d'Astrophysique de Paris, 98bis
  Boulevard Arago, 75013 Paris, France}
\email{gd@poly.in2p3.fr}

\begin{abstract}
The population synthesis of cataclysmic variable binary systems at
short orbital periods ($< 2.75$\,hrs) is investigated.  A grid of
detailed binary evolutionary sequences has been calculated and
included in the simulations to take account of additional angular 
momentum losses beyond that associated with gravitational radiation 
and mass loss, due to nova outbursts, from the system. As a 
specific example, we consider the effect of a circumbinary disk 
to gain insight into the ingredients necessary to reproduce the 
observed orbital period distribution.  The resulting distributions 
show that the period minimum lies at about
80\,minutes with the number of systems monotonically increasing with
increasing orbital period to a maximum near 90\,minutes. There is no
evidence for an accumulation of systems at the period minimum which is
a common feature of simulations in which only gravitational radiation
losses are considered. The shift of the peak to about 90\,minutes is a
direct result of the inclusion of systems formed within the period
gap.  The period distribution is found to be fairly flat for orbital
periods ranging from about 85 to 120\,minutes. 
The steepness of the lower edge of the period gap can be
reproduced, for example, by an input of systems at periods near
2.25\,hrs due to a flow of cataclysmic variable binary systems from
orbital periods longer than 2.75\,hrs.

The good agreement with the cumulated distribution function of
observed systems within the framework of our model indicates that the
angular momentum loss by a circumbinary disk or a mechanism which 
mimics its features coupled with a weighting
factor to account for selection effects in the discovery of such
systems and a flow of systems from above the period gap to below the
period gap are important ingredients for understanding the overall
period distribution of cataclysmic variable binary systems.
\end{abstract}


\section{Introduction}
Amongst the observed properties of cataclysmic variable binary systems
(CVs), the most reliably determined quantity is the orbital
period. The distribution of these orbital periods is fundamental, and
theoretical interpretation of its features requires knowledge of the
formation and evolution of these systems. Despite the increase in the
number of CVs with determined orbital period to more than 500
systems as a result of their discovery in optical surveys in the last
decade, their orbital period distribution has continued to defy
explanation.  This has, in part, been due to our poor knowledge of the
theoretical description of the angular momentum loss processes which
are essential for understanding the evolution of these systems.

The major uncertainty stems from our lack of a fundamental theory for
angular momentum loss associated with a magnetically coupled stellar
wind. The form of magnetic braking that has been the basis of most
previous works on the evolution of CVs traces back to the pioneering
work of Skumanich (1972), who determined that the stellar rotational
velocity of slowly rotating main sequence G type stars varied
inversely proportional to the square root of its age. Donor stars in 
CVs, however, typically rotate much more rapidly (with rotation periods 
$\la 6-7$\,hrs) than single stars of the same spectral type, and 
the implicit assumption that the form of angular momentum loss derived 
for slow rotation can be extrapolated and applied to fast 
rotators is without substantiation.  Based on theoretical grounds, 
a departure from this form is expected since 
the acceleration of the stellar wind changes from thermally driven 
to centrifugally driven as the rotation speed increases.  It is 
now generally accepted that the angular momentum loss rate from 
rapidly rotating single stars is severely overestimated
by the rate based on the use of the Skumanich relation (MacGregor \&
Brenner 1991; St\c{e}pie\'n 1995; Andronov et al. 2003; Ivanova \&
Taam 2003).

Other properties of the angular momentum loss associated with magnetic
braking have been inferred from the existence of a gap from 2.25\,hrs
to 2.75\,hrs in the observed orbital period distribution of CVs.  This
feature in the distribution has been interpreted in terms of a sudden
decrease in the angular momentum loss rate attributed to disrupted
magnetic braking (see Spruit \& Ritter 1983; Rappaport, Verbunt, \&
Joss 1983), possibly reflecting a transition of the donor star from
one with a radiative core and a convective envelope to one with a
fully convective structure.  The observational evidence supporting 
such a discontinuous change at the fully convective transition on the 
main sequence is, however, lacking (see for example, Andronov
et al. 2003).
     
In addition to the observationally unsubstantiated phenomenological
description of angular momentum loss related to the period gap (Kolb
2002), difficulties also exist in reproducing the high value of the
period minimum at 80\,minutes, the lack of accumulation of systems at
the period minimum predicted by the standard model, the almost
equal number of CVs above the upper edge of the gap at 2.75\,hrs 
and below the lower edge of the gap at 2.25\,hrs (e.g. Barker \& Kolb 
2003), and the spread of mass-transfer rates at a given orbital 
period (see Spruit \& Taam 2001).
     
Based on the work of Kolb \& Baraffe (1999), the period minimum could
be shifted to the observed value provided that additional angular
momentum loss processes operate such that the rate is increased by
about a factor of 4 above that due to gravitational radiation alone.
However, such a solution does not lead to a smearing out of the 
number of systems at the period minimum.

Other indirect evidence which suggests the need for additional angular 
momentum losses has been provided by Patterson (2001).  Specifically, 
by using a statistical argument that the average white dwarf mass 
is $0.7 M_{\odot}$, he deduced that there appears to be a systematic shift in 
the mass-radius relation for secondaries in short-period CVs, finding radii 
that are larger than those predicted from standard stellar models subjected 
to mass loss driven by gravitational radiation. Since the stellar radius 
is a function of the degree to which the donor is driven out of thermal 
equilibrium, higher angular momentum losses above that given by gravitational 
radiation are required. 

Against this backdrop, additional angular momentum losses associated 
with some form of residual magnetic braking or mass loss have been 
hypothesized. Although this is possible, it is not a viable solution 
unless it is different for CVs with different present or past system 
properties (e.g. different initial donor masses), for otherwise the
period spike at the period minimum would remain.  Further evidence
contrary to this hypothesis stems from theoretical studies suggesting
that magnetic braking is suppressed in magnetic CVs (see Li, Wu, \&
Wickramasinghe 1994). Observational evidence for the suppression of
magnetic braking in magnetic CVs results from the measurements of the
effective temperatures of magnetic white dwarfs (WDs) (Araujo-Betancor
et al. 2005; G\"ansicke \& Townsley 2005) showing that the magnetic
CVs are characterized by lower mass accretion rates than dwarf novae
at the same orbital period. Thus, even if residual braking is present,
it should be even less for the magnetic CVs, which, barring a
fortuitous coincidence, would conflict with the observational result
that the orbital period distributions of magnetic and non magnetic CVs
below the period gap are similar.

Among other possible mechanisms, consequential angular momentum losses 
associated with mass transfer from the donor have been explored by Barker 
\& Kolb (2003). By applying a generic prescription for the process, Barker 
\& Kolb (2003) find that the shift of the period minimum to the observed 
value could not be achieved even assuming the mechanism to be maximally 
efficient. 
 
Given that these alternative angular momentum loss mechanisms fail to
reproduce the observed short period distribution we seek an
alternative or additional agent to remove angular momentum from the
system.  As an example, we explore the consequences of CV evolution
based on the inclusion of a circumbinary (CB) disk (Spruit \& Taam
2001; Taam \& Spruit 2001) on the period distribution. Direct
detection of this matter by absorption line studies (Belle et
al. 2004) and infrared continuum studies (Dubus et al. 2004) has so
far been elusive and difficult to interpret, partly because of the
lack of accurate disk atmosphere models. By including a CB disk in the
evolution we not only provide an indication of the ingredients
necessary for an angular momentum loss process to reproduce the
observed period distribution, but also make it possible to identify
the basic properties of such circumbinary material in typical systems.

In order to demonstrate its possible effect on the
evolution of CVs, a population synthesis differential analysis is
required to examine the differences between the standard model and a
model invoking CB disks.  In this way, the strengths and weaknesses of
each can be identified to determine the properties of the angular
momentum loss processes that best reproduce the observed period
distribution.  In comparing the theoretical distributions with the
observed distributions, a treatment of the selection effects will be 
necessary since complete surveys of CVs to a given distance are not 
available.

In this paper, we report on the results of a population synthesis
for CVs formed at short orbital periods (less than 2.75\,hrs). In the
next section we describe the computational technique and outline the
assumptions inherent in this study.  The construction of the zero age
CV (ZACV) population will be presented in \S 3. The intrinsic period
distribution of CVs, a description of the selection effects, and their
application to the intrinsic orbital period distribution is presented
in \S 4.  Finally, we discuss the implications of our numerical
results and conclude in the last section.
 
\section{Computational technique}
\label{tech}

A hybrid binary population synthesis technique is used in which the
output of a rapid binary population synthesis code is combined with
detailed evolutionary tracks describing the evolution of CVs. The
approach is similar to that adopted by Kolb (1993) who, for
computational efficiency, combined secular evolution tracks from a
simplifying bipolytrope stellar evolution code with CV birth rates
computed by Politano (1988) and de Kool (1992).  We note that 
bipolytropes were also used by Howell, Nelson, \& Rappaport (2001) 
in a study that focused on the 2-3\,hr period gap in the observed 
CV orbital period distribution.  The main ingredients of the population 
synthesis scheme adopted in the present paper are described in more 
detail in the following subsections.

\subsection{The population synthesis code}

\subsubsection{Initial binary parameters}

The BiSEPS binary population synthesis code described by Willems \&
Kolb (2002, 2004) is used to construct a population of ZACVs in the
four-dimensional $\left( t, M_{\rm WD}, M_{\rm d}, P_{\rm orb}
\right)$-space, where $t$ denotes the time, $M_{\rm WD}$ the mass of
the WD, $M_{\rm d}$ the mass of the donor star, and $P_{\rm orb}$ the
orbital period. The initial primary and secondary masses are chosen
from a grid of 80 logarithmically spaced masses in the interval from
$0.1$ to $20\,M_\odot$, and the initial orbital periods from
a grid of 300 logarithmically spaced periods ranging from 0.1 to
10\,000 days. For symmetry reasons only binaries with $M_1 > M_2$ are
evolved. Our sample of zero-age main-sequence binaries then typically
consists of $\sim 10^6$ systems. All stars are assumed to have
Population~I chemical compositions. 

The probability that the primary (the WD progenitor) has an initial
mass $M_1$ is determined by the initial mass function
  \begin{equation}
  \renewcommand{\arraystretch}{1.4} \xi \left(M_1 \right) = \left\{
    \begin{array}{ll}
    0 & \hspace{0.3cm} M_1/M_\odot < 0.1, \\
    0.29056\, M_1^{-1.3}   & \hspace{0.3cm} 0.1 \le M_1/M_\odot < 0.5, \\
    0.15571\, M_1^{-2.2}   & \hspace{0.3cm} 0.5 \le M_1/M_\odot < 1.0, \\
    0.15571\, M_1^{-2.7} & \hspace{0.3cm} 1.0 \le M_1/M_\odot < \infty,
    \end{array}
  \right. \label{imf}
  \end{equation}
while the probability that the secondary (the future CV donor star)
has an initial mass $M_2$ is determined from the initial mass ratio
distribution
  \begin{equation}
  \renewcommand{\arraystretch}{1.4} n(q) = \left\{
    \begin{array}{ll}
    \mu\,q^\nu & \hspace{0.3cm} 0 < q \le 1, \\
    0 & \hspace{0.3cm} q > 1
    \end{array}
  \right. \label{imrd}
  \end{equation}
(Kroupa, Tout, \& Gilmore 1993; Hurley, Tout, \& Pols 2002).  In
Eq.~(\ref{imrd}), $q=M_2/M_1$ is the mass ratio of the stars when the
binary is formed, $\nu$ is a constant, and $\mu$ is a normalisation
factor depending on $\nu$. In addition to the initial mass ratio
distribution given by Eq.~(\ref{imrd}), we also consider the
possibility that the initial secondary mass is distributed
independently according to the same initial mass function as
$M_1$. The flat initial mass ratio distribution corresponding to
$\nu=0$ and $\mu=1$ is considered to be our reference
distribution. The distribution of initial orbital separations $a$,
finally, is assumed to be logarithmically flat between 3 and
$10^6\,R_\odot$, i.e.
  \begin{equation}
  \renewcommand{\arraystretch}{1.4} \chi (a) = \left\{
    \begin{array}{ll}
    0 & a/R_\odot < 3 \mbox{ or } a/R_\odot > 10^6, \\
    0.078636\, a^{-1} & 3 \le a/R_\odot \le 10^6
    \end{array}
  \right. \label{iosd}
  \end{equation}
(Iben \& Tutukov 1984, Hurley et al. 2002).

In order to derive absolute formation rates and numbers of systems
presently occupying the Galaxy, the above distribution functions,
which are all normalized to unity, need to be supplemented with a star
formation rate and a binary fraction. For this purpose, it is assumed
that all stars are in binaries, and that the Galaxy has an age of
10\,Gyr.  The star formation rate is then taken to be constant
throughout the life time of the Galaxy and normalized so that one
binary with $M_1 > 0.8\,M_\odot$ is born each year. This is in
agreement with the observationally inferred formation rate of WDs in
the Galaxy (Weidemann 1990).

\subsubsection{Binary evolution aspects}
\label{binev}

One of the critical evolutionary phases in the formation of CVs is the
common-envelope (CE) phase that leads to the formation of the WD.  For
ease of comparison, this phase is modelled in a manner similar to
other investigations by equating the binding energy of the envelope to
the change in orbital energy as
\begin{equation}
{{G \left( M_{\rm c} + M_{\rm e} \right) M_{\rm e}}
  \over {\lambda_{\rm CE}\, R_{\rm L}}} =
  \alpha_{\rm CE} \left[ {{G\, M_{\rm c}\, M_2} \over {2\, a_{\rm f}}}
  - {{G \left( M_{\rm c} + M_{\rm e} \right) M_2}
  \over {2\, a_{\rm i}}} \right]\!  \label{ce}
\end{equation}
(Webbink 1984). In this equation, $G$ is the gravitational constant,
$M_{\rm c}$ and $M_{\rm e}$ are the core and envelope mass of the
Roche-lobe filling star (the WD progenitor), $R_{\rm L}$ is the radius
of its Roche lobe, $M_2$ is the mass of the companion (the future CV
donor star), $a_{\rm i}$ and $a_{\rm f}$ are the orbital separations
of the binary at the start and at the end of the common-envelope
phase, $\lambda_{\rm CE}$ is a dimensionless structure parameter
determining the binding energy of the envelope, and $\alpha_{\rm CE}$
is the fraction of the orbital energy that is transferred to the
envelope. In our standard population synthesis model, we set the
product $\alpha_{\rm CE}\, \lambda_{\rm CE}=0.5$ (cf. de Kool 1992,
Politano 1996).

If a binary survives the CE phase, its ability to evolve into a CV
depends on the post-CE orbital separation. For sufficiently close
orbits, the binary may evolve into a semi-detached state either
through the nuclear expansion of the main-sequence secondary or
through the loss of orbital angular momentum via magnetic braking
and/or gravitational radiation.  Depending on the masses of the binary
components and the evolutionary state of the donor star, the ensuing
mass-transfer phase may initially take place on the thermal time scale
of the donor star.  In this study, we leave aside these more
complicated systems and focus on CVs whose formation is driven by
angular momentum losses from the system rather than the nuclear
evolution of the donor. For an elaborate discussion of CVs forming
through a thermal time scale mass-transfer phase, see Kolb \& Willems
(2005).

Due to the uncertainties in the magnitude and form of the angular
momentum loss rate associated with magnetic braking (see \S 1), we
focus on the CVs below the upper edge of the period gap and adopt a
simple prescription in our calculations assuming that it takes place
on a constant time scale of 10\,Gyr for main-sequence stars with
masses in the interval between $0.35\,M_\odot$ and $1.25\,M_\odot$.
For stars less massive than $0.35\,M_\odot$ and stars more massive
than $1.25\,M_ \odot$ we assume that magnetic braking is
ineffective. Orbital evolution due to gravitational radiation is
treated in the weak-field approximation using the formulae given by
Hurley et al. (2002). At the onset of the CV phase, the stability of
mass transfer is determined using the radius-mass exponents tabulated
by Hjellming (1987) and by assuming that mass transfer proceeds in the
fully non-conservative approximation. As mentioned above, we only
consider systems for which mass transfer is dynamically and thermally
stable.  The evolution of the binary after the onset of the CV phase
is followed using detailed evolutionary tracks obtained from a full
binary evolution code. The tracks and the main features of the code
are described in more detail in the following section.

\subsection{The evolutionary tracks}
\label{evtracks}

The input physics and the evolutionary code used in the construction
of the binary evolutionary sequences are described in Taam et al.
(2003).  They are based on an updated stellar evolution code
originally developed by Eggleton (1971, 1972). The stellar models were
characterized by a solar metallicity and a helium abundance of $Y =
0.26805$. As a simplifying approximation, the binary evolution is
assumed fully non-conservative as a result of efficient mass loss
during a nova explosion so that the mass of the WD is fixed throughout
the evolution. Based upon the work of Taam et al. (2003), the fraction
of the transferred mass corresponding to $10^{-5}$ was assumed to be
deposited into a CB disk.  The evolution of the CB disk was followed
using a fully implicit method (Dubus, Taam, \& Spruit 2002) and
implemented into the binary evolutionary calculations as described in
Taam et al. (2003). Hence, the angular momentum loss from binaries
with initial orbital periods less than about 2.75\,hrs includes
gravitational radiation losses, losses associated with mass ejections
in nova outbursts (with the lost material carrying a specific angular
momentum corresponding to the orbital motion of the WD), and losses
due to gravitational torques from a CB disk (Spruit \& Taam 2001).

For comparison with population synthesis studies based on evolutionary
tracks without CB disks, we have also constructed a second set of
binary sequences with angular momentum losses associated with
gravitational radiation and long-term mean mass-loss rates from
the system  due to nova explosions only.  

As we shall see in \S 4, the essential features which significantly 
influence the orbital period distributions result from the fact that the 
evolutionary tracks with CB disks do not converge to a common track 
near the period minimum (Taam et al.  2003).  
That is, the addition of this form of angular momentum loss leads to 
tracks which are a function of the initial properties of the binary system 
at the time at which Roche lobe overflow is initiated (see \S 4.2).

A detailed study of the orbital period distribution of CVs below the
period gap requires high resolution histograms with bin sizes of the
order of a few minutes. Since the grid of evolutionary sequences was
calculated with steps of $0.1\,M_\odot$ in the initial mass of the
donor star, corresponding to steps of $\simeq 60$\,min in the initial
period, the density of the available sequences was increased using
cubic spline interpolations between sets of neighboring tracks. In
both cases (evolution with or without CB disk), this increased the
number of binary evolution sequences for CVs forming with periods
below 2.75\,hrs from $\sim 50$ to $\sim 1000$.

\section{The ZACV population}

\subsection{Birth rates as functions of the orbital period}

To understand the systematic behavior of the orbital period
distributions resulting from the binary population synthesis
calculations, we first scrutinize the dependency of CV {\em formation}
on the model parameters. For this purpose, we compare the populations
of newborn CVs obtained by varying the product $\alpha_{\rm CE}\,
\lambda_{\rm CE}$ parameterizing the CE phase as well as the initial
mass ratio or the initial secondary mass distribution. In our
reference model, model A, we set $\alpha_{\rm CE}\, \lambda_{\rm CE} =
0.5$. In models~CE1 and~CE8, we respectively decrease and increase the
product $\alpha_{\rm CE}\, \lambda_{\rm CE}$ by a factor of 5,
independent of the evolutionary stage and envelope mass of the binary
component initiating the CE phase. In model DCE1, we assume the
envelope ejection process to be easier or more efficient for
dynamically unstable case~C mass transfer than for dynamically
unstable case~B mass transfer, as suggested by Fig. 1 in Dewi \&
Tauris (2000).  In our last model, model DCE5, we assume that low-mass
envelopes are easier to eject than high-mass envelopes based on the
importance of spin up of the envelope (see Sandquist et al. 2000). The
values of the $\alpha_{\rm CE}\, \lambda_{\rm CE}$ parameters adopted
in the different population synthesis models are summarised in
Table~\ref{models}. The range of values and parametrizations as well
as the adopted initial mass ratio or initial secondary mass
distributions were chosen to be sufficiently wide to emphasize the
differences in the resulting CV populations.

\begin{figure*}
\rotatebox{90}{\resizebox{6.0cm}{!}{\includegraphics{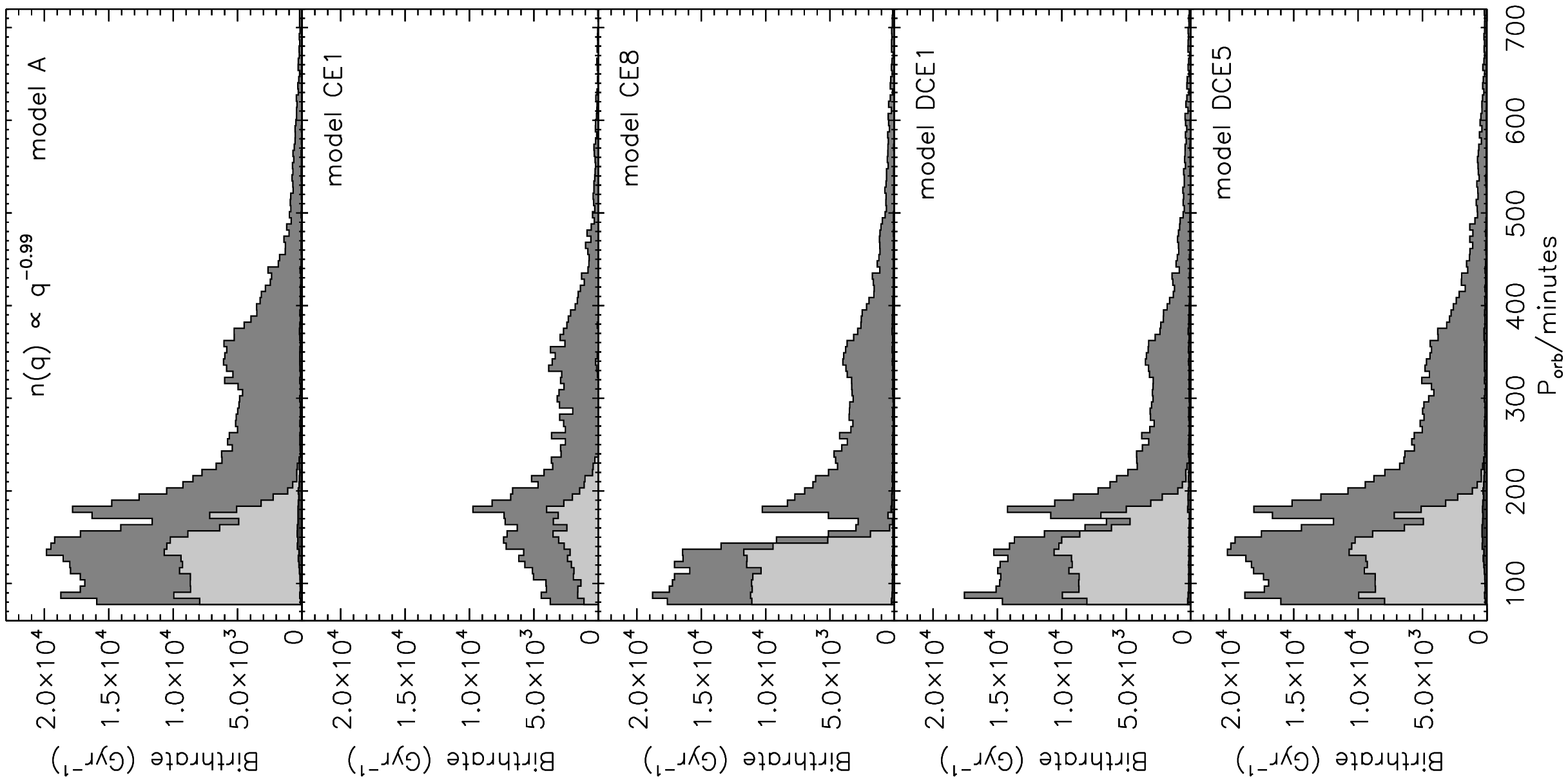}}
\resizebox{6.0cm}{!}{\includegraphics{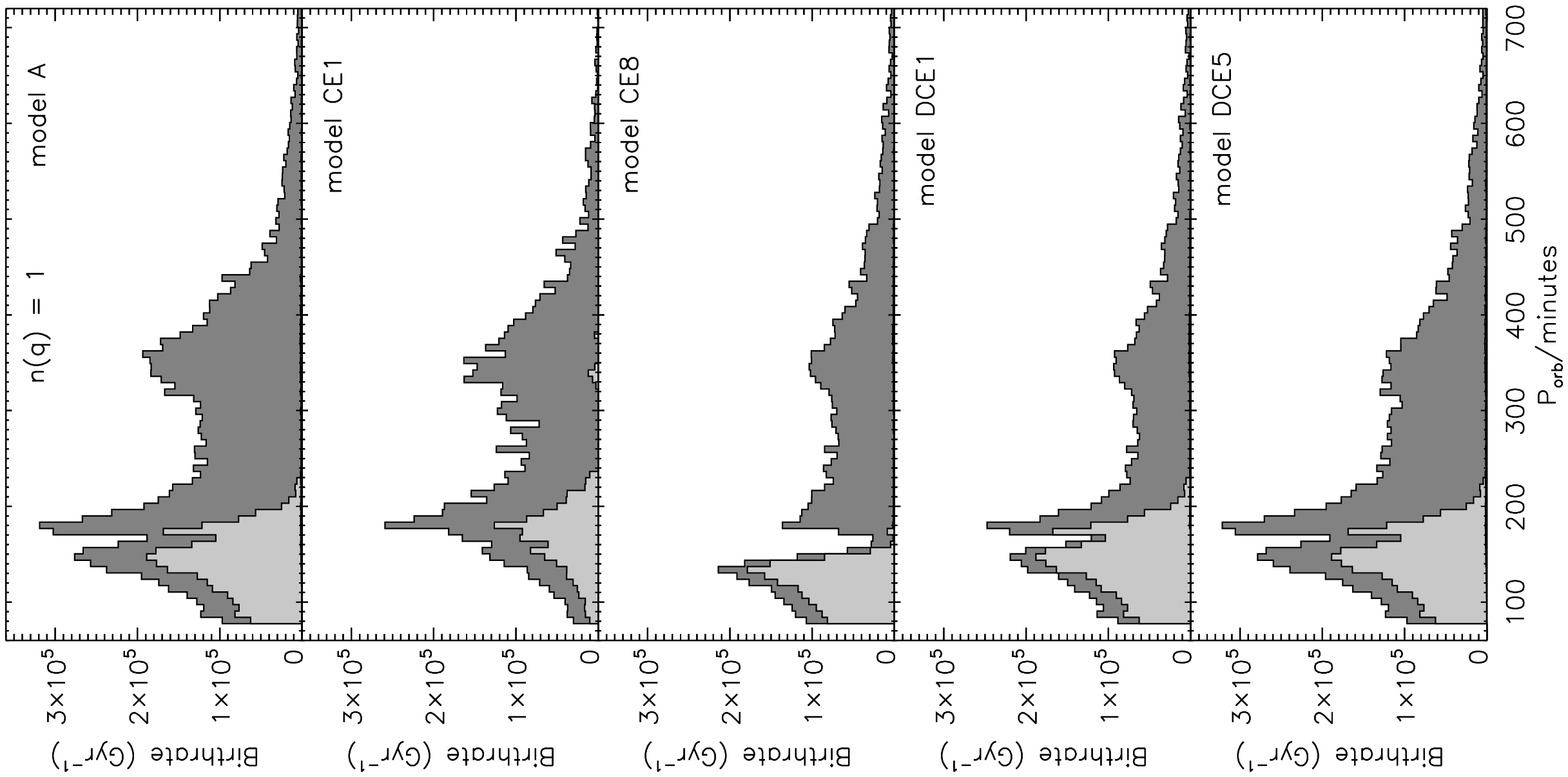}}
\resizebox{6.0cm}{!}{\includegraphics{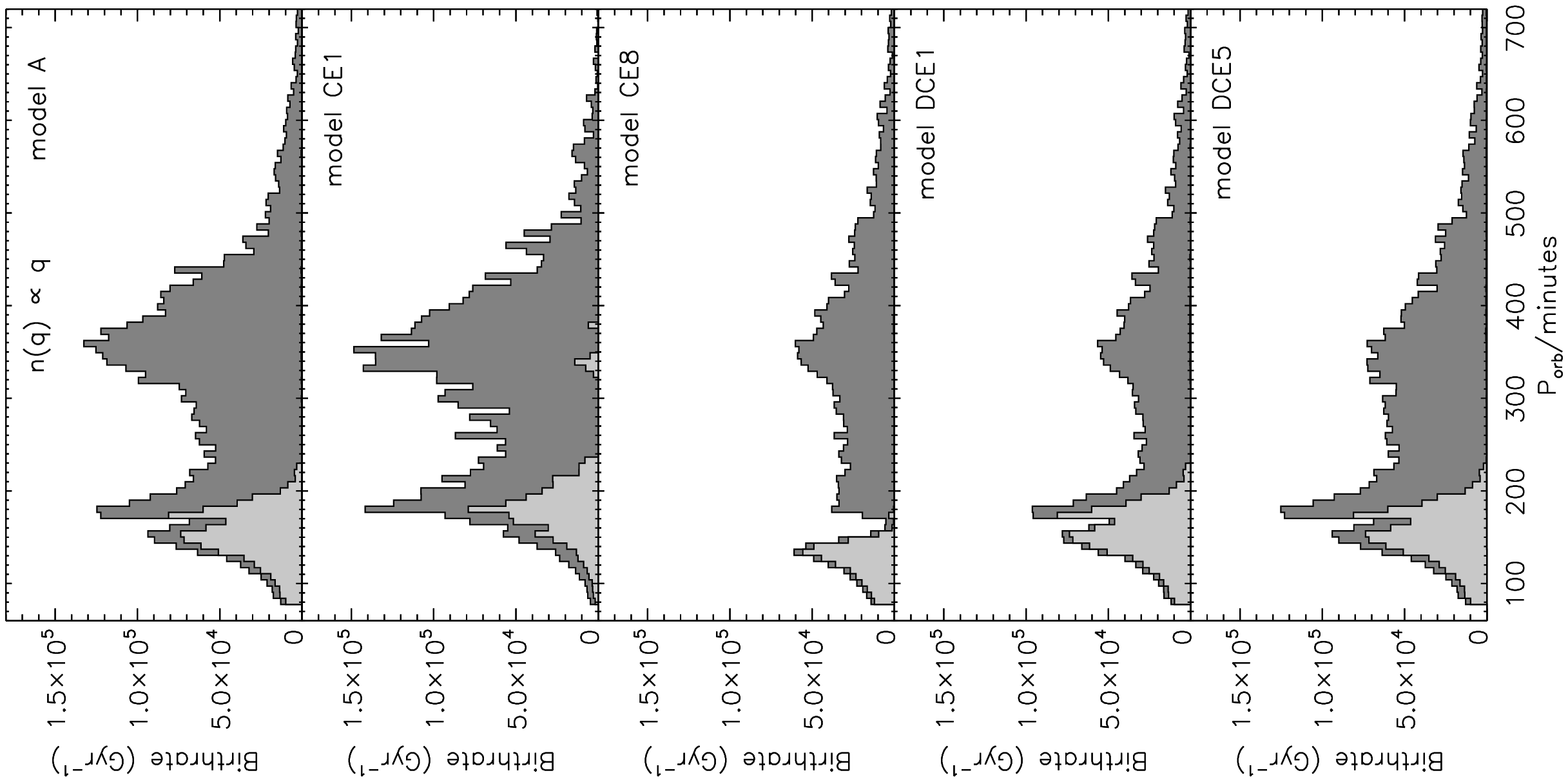}}}
\caption{\small Present-day formation rate of CVs in the Galactic disk
for different initial mass ratio distributions and different CE
models. The light and intermediate gray shading represent the
fractions of He and C/O white dwarf systems contributing to the
population. The contribution from O/Ne/Mg white dwarf systems is
negligible and therefore invisible on the scale of the figure.}
\label{zacv}
\end{figure*}

Fig.~\ref{zacv} shows the present-day CV formation rate as a function
of the orbital period for three different initial mass-ratio
distributions $n(q)=1$, $n(q) \propto q$, and $n(q) \propto
q^{-0.99}$. We note that we use $n(q) \propto q^{-0.99}$ instead of 
$n(q) \propto q^{-1}$ because the latter poses a normalization problem 
for mass ratios in the range $0 \le q \le 1$.

We first turn our attention to the case of the flat
initial mass-ratio distributions $n(q)=1$ and our reference model~A.
Going from longer to shorter orbital periods, the orbital period
distribution gently rises until it reaches a local maximum at $P_{\rm
orb} \simeq 350\, {\rm min}$. The following dip between $\simeq 220$
and $\simeq 320$\,min is associated with the dynamical instability of
mass transfer from donor stars with deep convective envelopes (see de
Kool 1992 for details). The steep rise in the birthrates towards even
shorter orbital periods is terminated by an abrupt decrease in the
number of CVs formed at periods just below $\simeq 180\, {\rm min}$
which, for ZAMS stars, corresponds to a donor mass $M_2 \approx
0.35\,M_\odot$. This decrease is associated with the switch-off of
magnetic braking for fully convective stars. Below $P_{\rm orb} \simeq
180\, {\rm min}$, gravitational radiation is therefore the only
angular-momentum loss mechanism available to bring white dwarf
main-sequence star binaries into contact. The local maximum at $\simeq
140-150$\,min is due to CVs containing He WDs which, because of mass
transfer stability requirements, are absent at orbital periods longer
than $\simeq 200$\,min.

From Eq.~(\ref{ce}), it is easily seen that the post-CE orbital
separations increase with increasing values of $\alpha_{\rm CE}\,
\lambda_{\rm CE}$. For case B RLO systems this implies that the number
of systems surviving the CE phase increases with increasing
$\alpha_{\rm CE}\, \lambda_{\rm CE}$. Hence, for larger values of
$\alpha_{\rm CE}\, \lambda_{\rm CE}$, more CVs with He WDs are
formed. For case C RLO systems, the orbital separation at the onset of
the CE phase is larger, so that survival of the CE phase is less of an
issue. However, increasing $\alpha_{\rm CE}\, \lambda_{\rm CE}$ makes
it more difficult for WD systems with main sequence companions to
evolve into contact after the CE phase. Consequently, the ratio of the
number of systems forming with He WDs (at periods below $\simeq
200$\,min) to the number of systems forming with C/O and O/Ne/Mg WDs
(mainly at periods above $\simeq 200$\,min) increases with increasing
$\alpha_{\rm CE}\, \lambda_{\rm CE}$. These dependencies explain the
basic differences between the orbital period distributions found for
different CE model parameters (see Table~\ref{models}). The
distribution of the formation rates of systems with He WDs is
furthermore seen to be particularly sensitive to the CE model. For
$\alpha_{\rm CE}\, \lambda_{\rm CE}=2.5$ (model~CE8), we note a
pronounced gap in the ZACV orbital period distribution near $P_{\rm
orb} \simeq 150-180\,{\rm min}$, caused by a decrease of the upper
limit on the He WD mass in the CV population which in turn causes a
decrease in the maximum donor star mass leading to stable mass
transfer. This decrease is related to the fact that the most massive
He WDs stem from late case~B CE phases which, for a large CE ejection
efficiency, lead to post-CE orbits that are too wide for the system to
evolve into a CV within the adopted Galactic age limit of
10\,Gyr. Systems with O/Ne/Mg WDs have very small formation rates in
all models considered.

A mass ratio distribution $n(q) \propto q^{-0.99}$ favoring lower
$q$-values (or, for a given $M_1$, lower $M_2$-values) increases the
relative number of systems below $\simeq 160$\,min in all models
considered. A mass ratio distribution function such that the secondary
mass $M_2$ is chosen independently from the same IMF as $M_1$ leads to
a similar result in the relative number of systems below $\simeq
160$\,min with the effect slightly enhanced.  The shape of the orbital
period distribution is furthermore also very similar in these two
cases. In particular, the peak of the orbital period distribution at
$\simeq 350$\,min is in both cases strongly reduced to a tiny bump
that is barely visible in the distributions. On the other hand, a mass
ratio distribution $n(q) \propto q$ favoring equal mass ratios tends
to increase the relative number of systems forming at longer orbital
periods or, equivalently, the relative number of systems with higher
donor masses, leading to peaks of the period distribution of
comparable heights at $\simeq 180$\,min and $\simeq 350$\,min. We also
note that the relative contribution of He WD systems to the CV birth
rate at periods shorter than $\simeq 180\, {\rm min}$ is smallest for
$n(q) \propto q^{-0.99}$ and largest for $n(q) \propto q$. In the case
of the initial mass ratio distribution $n(q) \propto q$, our standard
model (model A) furthermore shows the same basic features as found by
de Kool (1992) and Politano (1996), even though their results were
derived under the assumption that the mass transfer process in the
system is conservative. The $n(q) \propto q^{-0.99}$ case can
furthermore be compared favorably to de Kool's (1992) results for an
independent choice of $M_2$ taken from the same IMF as $M_1$.

\subsection{Absolute and relative birth rates of CVs in the Galactic
  disk} 

Table~\ref{br} lists the total birth rate of all CVs forming at
periods shorter than 2.75\,hrs as well as the separate birth rates for
CVs containing He, C/O, and O/Ne/Mg WDs at the current epoch. For a
given initial mass ratio or secondary mass distribution, the total
birth rate of CVs below 2.75\,hrs changes by at most a factor of 2-3
between different population synthesis models. The birth rates are
largest when $M_2$ is distributed independently according to the same
IMF as $M_1$ (on the order of $10^{-2}\,{\rm yr^{-1}}$), and smallest
when $n(q) \propto q^{-0.99}$ (on the order of $10^{-4}\,{\rm
yr^{-1}}$). Consequently, even though the shape of the distribution
functions of the birth rate as a function of orbital period is similar
in these two cases, their magnitudes are not. In the case of a flat
initial mass ratio distribution $n(q)=1$, the birth rates are of the
order of $10^{-3}\,{\rm yr^{-1}}$.

When the birth rates of CVs with different types of WDs are considered
separately, the formation rates of systems with O/Ne/Mg WDs are always
2 to 3 orders of magnitude smaller than those of systems with He or
C/O WDs. This is a direct result of the rapid decrease of the initial
mass function with increasing mass $M_1$ of the WD progenitor as well
as the relatively small primary mass range giving rise to O/Ne/Mg WDs
($M_1 \simeq 7-8\,M_\odot$). In the case of population synthesis model
CE1, no systems with O/Ne/Mg WDs are formed below $\simeq 8$\,hrs due
to the decrease of the post-CE orbital separation with increasing mass
$M_1$ of the WD progenitor: when combined with the small CE ejection
efficiency $\alpha_{\rm CE}\, \lambda_{\rm CE}=0.1$, this behavior
requires large pre-CE orbital separations in order for the binary to
survive the CE phase. This leaves only a small range of separations
for which the secondary is still able to fill its Roche lobe while on
the main-sequence. Outside this range, Roche-lobe overflow from the
WD's companion does not occur until it becomes a giant star.

Depending on the initial mass ratio or initial secondary mass
distribution and the adopted CE model, the contribution of systems
with He WDs to the birth rate of Galactic CVs below 2.75\,hrs ranges
from 40 to 90\%. In the case of model~A and a flat initial mass ratio
distribution, about 65\% of all CVs below 2.75\,hrs are born with He
WDs. The relative contribution of systems with He WDs to the CV birth
rate is furthermore always largest for models CE8 and DCE1.
Correspondingly, the relative contribution of systems with C/O WDs is
smallest for these models. This is caused by the wider post-CE orbital
separations of the C/O WD CV progenitors which significantly reduces
the number of systems able to become semi-detached within the age of
the Galaxy. Similarly, the relative contribution of systems with He
WDs is smallest for model CE1 because the small $\alpha_{\rm CE}\,
\lambda_{\rm CE}$ value makes it harder for the progenitors of these
systems to survive the CE phase that forms the WD. Finally, for a
given CE model, the relative contribution of C/O WD systems to the CV
birth rate below 2.75\,hrs decreases from $n(q) \propto q^{-0.99}$ to
$n(q) \propto q$. This is caused by the fact that C/O WDs in CVs are
formed from stars with initial masses larger than about
$1.5\,M_\odot$. Hence, if initial mass ratios close to unity are
favored in the primordial population, the majority of the donor stars
in CVs containing C/O WDs will be too massive to initiate a
dynamically and thermally stable mass-transfer phase.

\section{The orbital period distribution below 2.75\,hrs}

In the following subsections, we focus on the theoretical present-day
orbital period distribution of CVs with periods below 2.75 hr, and
compare population synthesis results for systems evolving under the
influence of a CB disk and gravitational radiation with results for
systems evolving under the influence of gravitational radiation
only. In both cases, we adopt the fully non-conservative approximation
and, unless stated otherwise, assume that no systems formed at periods
longer than 2.75\,hrs evolve to periods below 2.75\,hrs. The
distributions are also compared with the observed CV orbital period
distribution consisting of nonmagnetic and magnetic CVs. For ease of
comparison, all distribution functions are normalised to unity. An
absolute scale can be obtained by multiplying the probability
distribution functions by the adopted bin size ($\simeq 2.8$\,min) and
the absolute numbers of systems listed in Table~\ref{num}.

The results of our population synthesis calculations are organized as
follows.  In \S 4.1, the intrinsic present-day CV population without
account of observational selection effects is examined. We note,
however, that a direct comparison of the theoretical and observed
orbital period distributions is only possible if a complete sample of
observed CVs is available. In \S 4.2, we therefore model the possible
selection effects in a simple manner by weighing the contribution of a
system to the population of CVs according to a factor determined by
the accretion luminosity. Detailed modeling of the selection effect is
highly uncertain since it depends, for example, on the bolometric
corrections of the emission from a non steady accretion disk, and it
is beyond the scope of the present investigation.  Hence, we adopt a
simple luminosity weighting factor to provide us with at least an
indication of the observational selection effects.  A summary of the
main theoretical results of \S 4.1 and \S 4.2 is provided at the end
of each subsection to facilitate understanding of the main physical
effects. The distribution of mass transfer rates and donor
masses as a function of orbital period is discussed in \S 4.3.
Finally, the typical CB disk properties of the present day CV population
are described in \S 4.4. 

\subsection{The intrinsic CV population}

\subsubsection{CVs evolving under the influence of gravitational
radiation and the long-term mean mass-loss rate from the system 
due to nova explosions} 

\begin{figure*}
\rotatebox{90}{\resizebox{9.0cm}{!}{\includegraphics{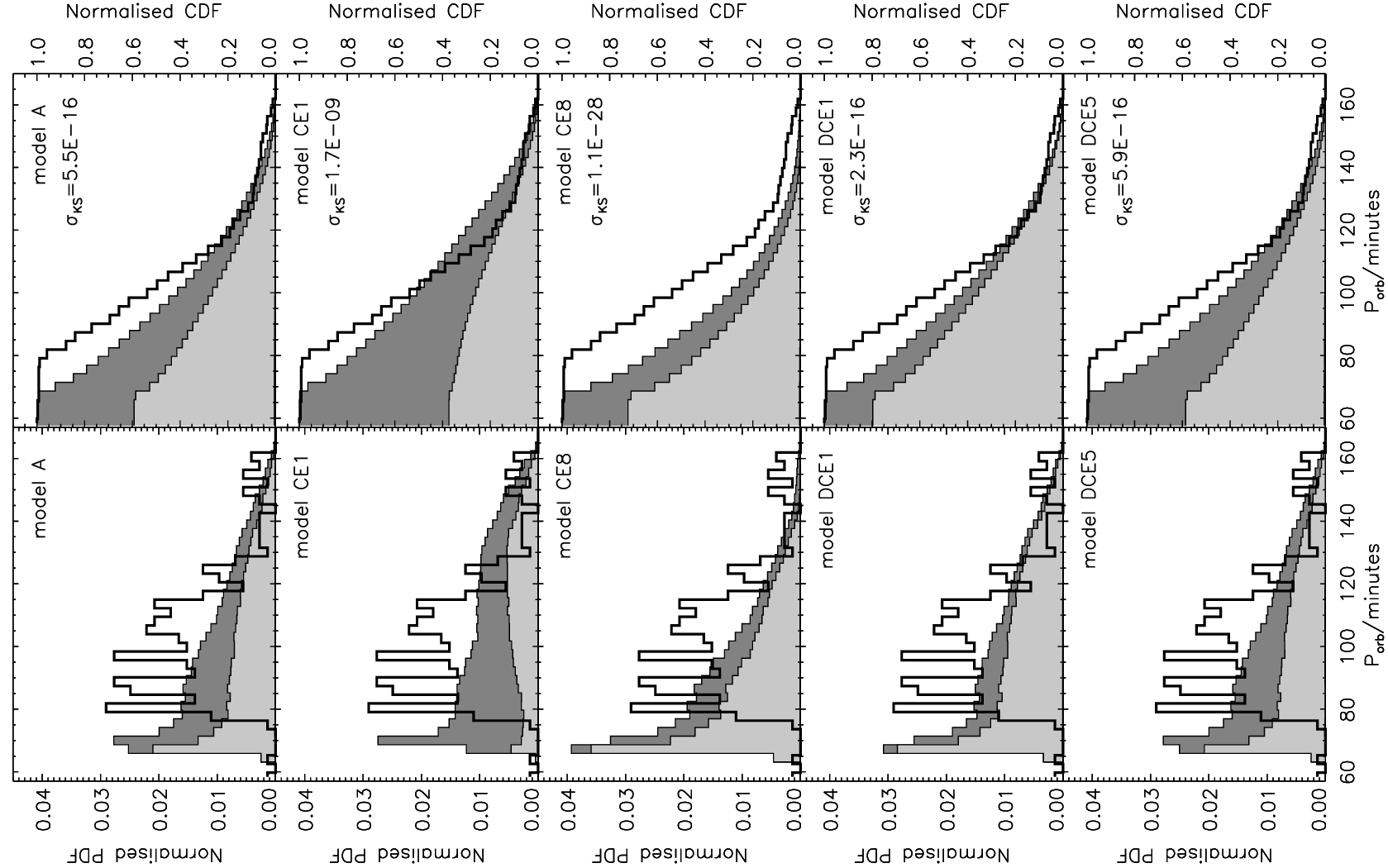}}
\resizebox{9.0cm}{!}{\includegraphics{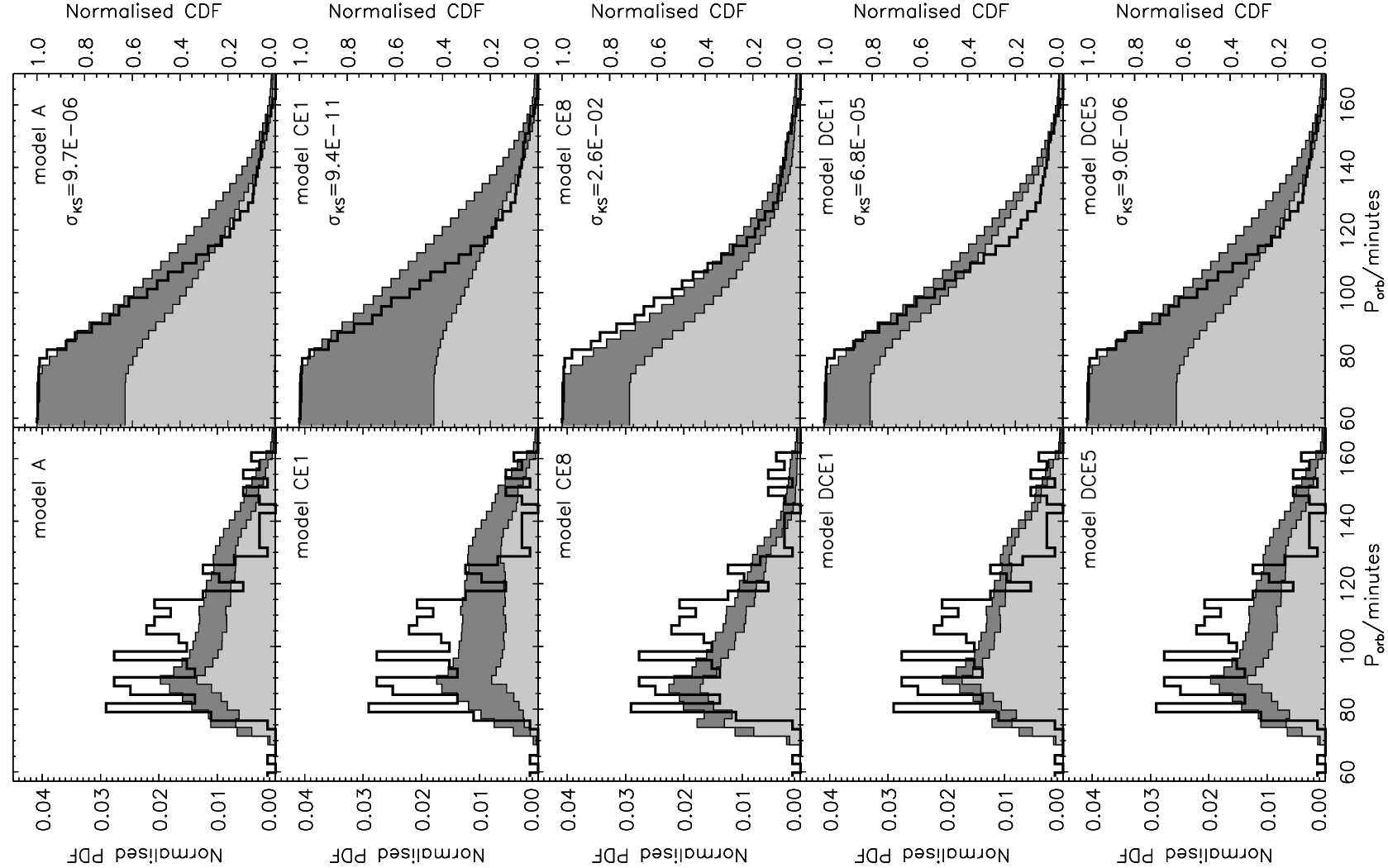}}}
\caption{\small Intrinsic orbital period distribution of Galactic CVs
for an initial mass ratio distribution $n(q)=1$ and different CE
models, without regard for observational bias. The left (right) most
panels are obtained from evolutionary tracks without (with) a CB
disk. The light and intermediate gray shading represent the fractions
of He and C/O white dwarf systems contributing to the population. The
contribution from O/Ne/Mg white dwarf systems is invisible on the
scale of the figure. The thick solid represents the observed CV
orbital period distribution. For ease of comparison, the PDFs and CDFs
are normalized to unity. The agreement between the observed and
simulated CDFs is indicated by the Kolmogorov-Smirnov significance
level $\sigma_{\rm KS}$.} 
\label{nodisk}
\end{figure*}

The intrinsic present-day probability distribution functions (PDFs)
and cumulative distribution functions (CDFs) of the orbital periods of
Galactic CVs based on evolutionary tracks with angular momentum losses
associated with gravitational radiation and the long-term mean
mass-loss rate from the system due to nova explosions are shown in
the left most panels of Fig.~\ref{nodisk} for an initial mass ratio
distribution characterized by $n(q)=1$.  Here the CDFs are defined to
be increasing with decreasing period below 2.75\,hrs (cf. Kolb
1995). For comparison, the orbital period distribution of observed CVs
obtained from the January 2005 edition (RKcat7.4) of the Ritter \&
Kolb (2003) catalog is represented by a thick solid line. The degree
of agreement or disagreement between the observed and simulated
orbital period distributions presented is quantified by means of a
Kolmogorov-Smirnov statistic and its associated significance level
$\sigma_{\rm KS}$ listed in the CDF-panels of the figure. We recall
that the latter is a measure of the probability that two data sets are
drawn from the same parent distribution. In particular, $\sigma_{\rm
KS}=0$ indicates that the CDFs of the orbital periods of observed and
simulated CV populations are significantly different, while
$\sigma_{\rm KS}=1$ indicates that the two CDFs are in very good
agreement.

Regardless of the adopted CE model, the PDFs always exhibit a sharp
increase in the number of systems near $\simeq 70$\,min which is
not seen in the observed distribution (see also Kolb \& Baraffe 1999,
Barker \& Kolb 2003). At longer periods all PDFs gently rise from
longer to shorter orbital periods until they reach the sharp peak near
the minimum period. In the case of model~CE1, the PDF exhibits a
plateau between $\simeq 100$ and $\simeq 130$\,min. The minimum period
in the theoretical distributions is furthermore systematically offset
with respect to the observed minimum period by about 10\,min. From the
CDFs, it is furthermore clear that the theoretical distributions
generally have a slight overabundance of systems above $\simeq
110-120$\,min, and a significant underabundance of systems between
$\simeq 80$ and $\simeq 110$\,min. The overabundance at longer orbital
periods is most severe for model~CE1. The slightly different behavior
of the orbital period distribution in model CE1 is due to the absence
of a gap between $\simeq 150$ and $\simeq 170$\,min in the birth rate
which is present in all other models (see Fig.~\ref{zacv}).

\begin{figure*}
\resizebox{8.5cm}{!}{\includegraphics{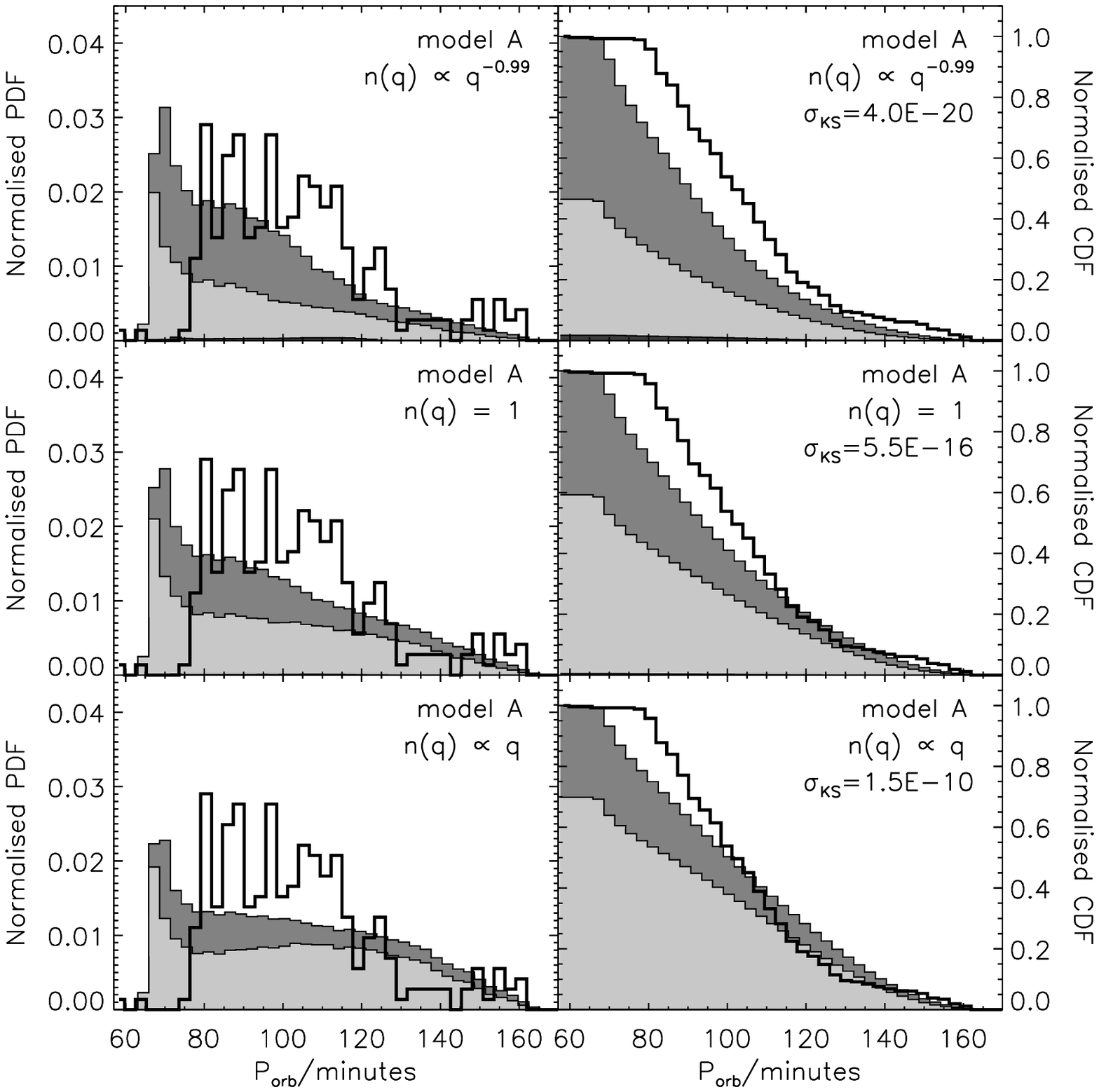}}
\resizebox{8.5cm}{!}{\includegraphics{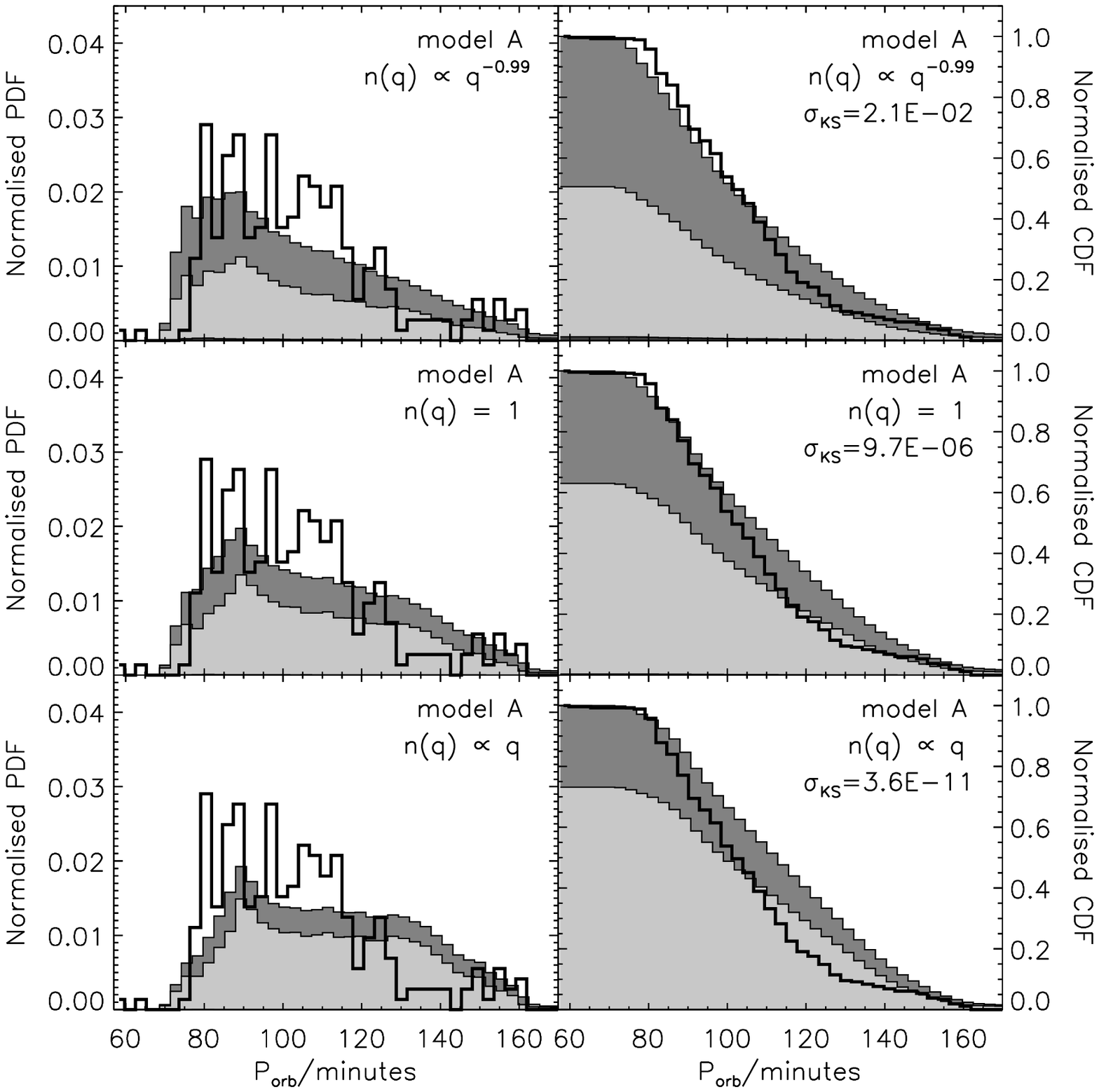}}
\caption{\small As Fig.~\ref{nodisk}, but for population synthesis
  model A and different initial mass ratio distributions.}
\label{nodisk2}
\end{figure*}

In the case of population synthesis model~A, 
the effect of the initial mass ratio distribution on the intrinsic
orbital period distribution is illustrated in the left most panels of
Fig.~\ref{nodisk2}. The
initial mass ratio distribution predominantly affects the relative
shape of the distribution functions in the region between $\simeq 90$
and $\simeq 140$\,min. Increasing the weight of systems with large
initial mass ratios depletes this region, while increasing the weight
of systems with equal mass ratios boosts this region. This
depletion/boost of systems is mainly caused by a decrease/increase in
the relative number of systems with He WDs. This result leads to a
flattening of the orbital period distribution between $\simeq 80$\,min
and $\simeq 120$\,min when $n(q) \propto q$.  The overabundance of
systems at periods above $\simeq 110$\,min furthermore becomes more
pronounced when $n(q) \propto q$. In the case of the initial mass
ratio distribution $n(q) \propto q^{-0.99}$, on the other hand, the
overabundance disappears completely and the theoretical CDF lays below
that of the observed systems over the entire range of orbital periods
considered. The shape of the PDF and CDF obtained for an independent
$M_2$ distribution according to the same IMF as $M_1$, is almost
indistinguishable from that obtained for $n(q) \propto q^{-0.99}$.

\subsubsection{CVs evolving under the influence of a CB disk}

The PDFs and CDFs resulting from the inclusion of a CB disk in the
evolutionary tracks used for the population synthesis simulations is
illustrated in the right most panels in Fig.~\ref{nodisk}.  As in the
left most panels, the PDFs and CDFs correspond to a flat initial mass
ratio distribution $n(q)=1$.  It can be seen that the introduction of
a CB disk lessens the degree of accumulation of systems near the
minimum period, resulting in a flattening of the orbital period
distribution in all five models. This behavior yields a significant
improvement of the CDFs at periods below $\simeq 100$\,min where there
is no longer a severe underabundance of systems with respect to the
observed CDF. In addition, the minimum period in the theoretical
distributions is now much closer to the minimum period in the observed
distribution. The Kolmogorov-Smirnov significance levels confirm the
large overall improvement in the agreement between the observed and
simulated orbital period distributions when the effects of a CB disk
are included.

As illustrated in the right most panels of Fig.~\ref{nodisk2}, varying
the initial mass ratio distribution has overall the same effect as for
the population synthesis simulations without a CB disk. For $n(q)
\propto q^{-0.99}$, the region between $\simeq 90$ and $\simeq
140$\,min gets depleted, resulting in a monotonic increasing PDF with
decreasing orbital periods from $\simeq 165$ to $\simeq 90$\,min. This
is in contrast and opposite to the case for the initial mass ratio
distribution $n(q) \propto q$ where the contribution of systems in
this region gets boosted and the PDFs are flattened even further. The
change in shape according to the adopted initial mass ratio
distribution is caused by the increase in the relative number of
systems with He WDs when equal initial mass ratios are favored. For
$n(q) \propto q^{-0.99}$, the moderate peak near $\simeq 90$\,min
which is present for $n(q)=1$ and $n(q) \propto q$ is furthermore
transformed into a local plateau ranging from $\simeq 75$ to $\simeq
90$\,min.  The PDFs and CDFs for $M_2$ distributed independently
according to the same IMF as $M_1$, are again very similar to those
obtained for $n(q) \propto q^{-0.99}$.

\subsubsection{Absolute and relative numbers of systems}

The total number of CVs with orbital periods shorter than 2.75\,hrs in
the Galactic disk as well as the relative number of systems containing
He, C/O, and O/Ne/Mg WDs are listed in Table~\ref{num}. For a given
initial mass ratio or initial secondary mass distribution, the total
number of systems typically changes by less than a factor of 3 between
different population synthesis models. The relative numbers of systems
with He, C/O, and O/Ne/Mg WDs, on the other hand, does change
significantly when different CE treatments are adopted. The relative
contribution of systems with He WDs to the population of CVs is always
smallest for model CE1 and largest for model DCE1. The relative number
of systems with C/O WDs shows the opposite trend. This is a direct
consequence of the dependence of the post-CE orbital separation on the
product $\alpha_{\rm CE}\, \lambda_{\rm CE}$ as described in \S
\ref{binev}. Systems with O/Ne/Mg WDs always provide a very small
contribution to the populations of CVs (less than 2\%) regardless of
the adopted CE model.

The absolute numbers of systems can change by as much of an order of
magnitude if different initial mass ratio or initial secondary mass
distributions are adopted. The total number of systems is largest
($\sim 10^8$) when $M_2$ is distributed independently according to the
same IMF as $M_1$ and smallest ($\sim 10^6$) when $n(q) \propto
q^{-0.99}$. In the case of a flat initial mass ratio distribution, the
total number of systems is of the order of $10^7$. The relative number
of systems with He WDs furthermore benefits from initial mass ratio
distributions favoring equal mass ratios and decreases by about 20\,\%
when more extreme mass ratios are favored or when $M_2$ is assumed to
be distributed independently. It is interesting to note that $n(q)
\propto q^{-0.99}$ and an independent $M_2$ distribution yield very
similar fractions of He, C/O, and O/Ne/Mg WD systems, but differ in
absolute numbers of systems by almost two orders of magnitude.

These results on the absolute and relative numbers of systems apply to
systems evolving under the influence of gravitational radiation and
the long-term mean mass-loss rate from the system due to nova
explosions only as well as to systems evolving under the influence of
a CB disk as well. The numbers of systems are furthermore strikingly
similar in both cases.  That is, the total number of systems is
systematically smaller when the effects of a CB disk are taken into
account, but the difference is always smaller than a factor of $\sim
2$. The slight decrease in the number of systems is caused by the
shorter life time of systems with CB disks in comparison to systems
evolving under the influence of gravitational radiation and the
long-term mean mass-loss rate from the system due to nova
explosions. The relative number of systems with He, C/O, and O/Ne/Mg
WDs are also similar although overall, there is a slight increase in
the fraction of systems with He WDs and a correspondingly small
decrease in the fractions of systems with C/O and O/Ne/Mg WDs when a
CB disk contributes to the evolution. The latter is caused by the
slightly more massive donor stars in C/O and O/Ne/Mg WD systems which
lead to somewhat more massive CB disks and thus a slightly faster
evolution.

\subsubsection{Pre-bounce vs. post-bounce systems}

We conclude the discussion of the present-day intrinsic CV
population with a comparison between the relative numbers of systems
that are evolving from long to short orbital periods (pre-bounce
systems) and the relative numbers of systems that are evolving from
short to long orbital periods (post-bounce systems). These relative
numbers as well as their decomposition according to the type of WD in
the system are listed in Table~\ref{prepost}. For a given initial mass
ratio or initial secondary mass distribution, the relative numbers of
pre- and post-bounce systems typically change by less than 15\%
between different population synthesis models. Changing the initial
mass ratio or initial secondary mass distribution, on the other hand,
introduces variations in the relative numbers of the order of
25\%. For a given population synthesis model, the relative number of
post-bounce systems is largest (30-45\%) when $n(q) \propto q^{-0.99}$
or when $M_2$ is distributed independently according to the same IMF
as $M_1$, and smallest (15-25\%) when $n(q) \propto q$. These
conclusions apply to systems evolving under the influence of
gravitational radiation as well as to systems evolving under the
influence of both gravitational radiation and a CB disk.

In the specific case of gravitational radiation, the
contributions of systems with He WDs and C/O WDs to the population of
post-bounce CVs tend to be within a factor of 2-3 from each other,
except for model~CE1 where the simulations predict a strong dominance
of systems with C/O WDs beyond the period minimum. However, this
dominance disappears completely when the effects of a CB disk on the
evolution of CVs are taken into account. It is also interesting to
note that in the case if the initial mass ratio distribution $n(q)
\propto q$, the total fraction of pre- and post-bounce systems is very
insensitive to whether or not the evolution is affected by a CB
disk. For $n(q) \propto q^{-0.99}$, on the other hand, the inclusion
of a CB disk typically decreases the total fraction of post-bounce
systems by about 10\%. This decrease is associated with the fact that
this mass ratio distribution favors systems with more massive donor
stars, and thus systems with more massive CB disks. The higher disk
mass in turn yields higher post-bounce mass-transfer rates and thus
shorter CV life times.

\subsubsection{Summary}

In summary, the theoretical intrinsic present-day orbital
distributions of CVs evolving under the influence of gravitational
radiation all show an accumulation of systems near a minimum period of  
$\simeq 65-70$\,min.  The value of this period minimum is in
approximate agreement with the results of Kolb \& Baraffe (1999)
and Barker \& Kolb (2003), but is longer by about $\simeq
5-10$\,min than that found in Kolb (1993), Kolb \& de Kool (1993), and
Howell et al. (2001).  This difference likely results from the
difference between evolutionary tracks calculated in this work,
Kolb \& Baraffe (1999), and Barker \& Kolb (2003), based on detailed
binary evolutionary tracks, as compared to bipolytropic models used in
Kolb (1993), Kolb \& de Kool (1993), and Howell et al. (2001).

The inclusion of an angular momentum loss associated with a CB disk
significantly reduces the accumulation of systems and increases the
minimum period to $\simeq 75-80$\,min. This lack of accumulation and
increase of the minimum period yield a significant improvement in the
modelling of the observed orbital period distribution of Galactic
CVs. The shift to a longer minimum period is a direct result of the
higher mass transfer rates associated with an additional angular
momentum loss process, whereas the smearing out of systems near the
period minimum is a consequence of the fact that the evolutionary
sequences do not converge to a common track.

Regardless of whether or not a CB disk is included in the CV
evolution, our population synthesis models predict a present-day
Galactic CV population consisting of $10^6$-$10^8$ systems, 15-45\% of
which have evolved beyond their minimum attainable period. The largest
uncertainties in these numbers stem from the uncertainties in the
initial mass ratio or initial secondary mass distribution.

\subsection{The observed CV population}

\subsubsection{Observational selection effects}

In order to properly compare the theoretical orbital period
distributions with the observed one, account must be taken of
observational selection effects. Since the discovery probability
depends on the flux from the source, one must apply a weighting factor
to the intrinsic distribution. Due to the low mass-transfer rates, the
population of CVs below the period gap is dominated by transients. In
the absence of a satisfactory theory of accretion disk outbursts that
would allow us to predict the outburst magnitude and recurrence time
as a function of the system parameters, systems in outburst and in 
quiescence are not distinguished. This assumption particularly neglects 
the effects of potentially long recurrence timescales as in WZ Sge. We 
furthermore assume that the sources are distributed isotropically within 
the Galactic disk such that the distance to which the sources are sampled 
is proportional to $L_{\rm acc}^{1/2}$, where the accretion luminosity is 
determined by $L_{\rm acc}=GM_{\rm WD} \dot{M_{\rm d}}/R_{\rm WD}$
(with $\dot{M_{\rm d}}$ the mean secular mass-transfer rate, and
$R_{\rm WD}$ is the radius of the WD). Since the mean mass-transfer
rates below the period gap are low (of order of $10^{-10} M_\odot {\rm
yr^{-1}}$) in both the evolutionary sequences with and without a CB
disk, their sampled distance is expected to be small compared to the
scale height of the galactic disk, so that the total observable
volume scales as $L_{\rm acc}^{3/2}$.  Hence, we model the
observational selection effects by subjecting the simulated
populations to a weighting factor, $W$, equal to $L_{\rm acc}^{3/2}$
(cf. Kolb 1993). The general effect of the dependence of the weighting
factor on the WD mass and radius is to favor system with more massive
WDs\footnote{Previous investigations (e.g. Howell et al. 2001) often
adopted a simple weighting factor proportional to $\dot{M_{\rm
d}}^{3/2}$ which neglects the dependence on the WD mass and radius. As
a test, we therefore ran several models using an $\dot{M_{\rm
d}}^{3/2}$ and found the differences with the results obtained for the
$L_{\rm acc}^{3/2}$ to be small.}.

We furthermore note that the observed distribution function of
magnetic CVs is similar to that of the non magnetic CVs, suggesting
that a similar weighting factor might be appropriate for both systems
in the lowest order of approximation. Although different weighting
factors may be considered for these two subpopulations (with
corresponding different population synthesis model parameters), it
would seem that some fine tuning would be necessary to fit to the
observed distribution. To avoid the additional complications
associated with different weighting factors (e.g., different 
bolometric corrections) of these two populations, we adopt the same 
weighting factor for both populations and treat the populations together. 

\subsubsection{CVs evolving under the influence of gravitational
  radiation only}

\begin{figure*}
\resizebox{8.5cm}{!}{\includegraphics{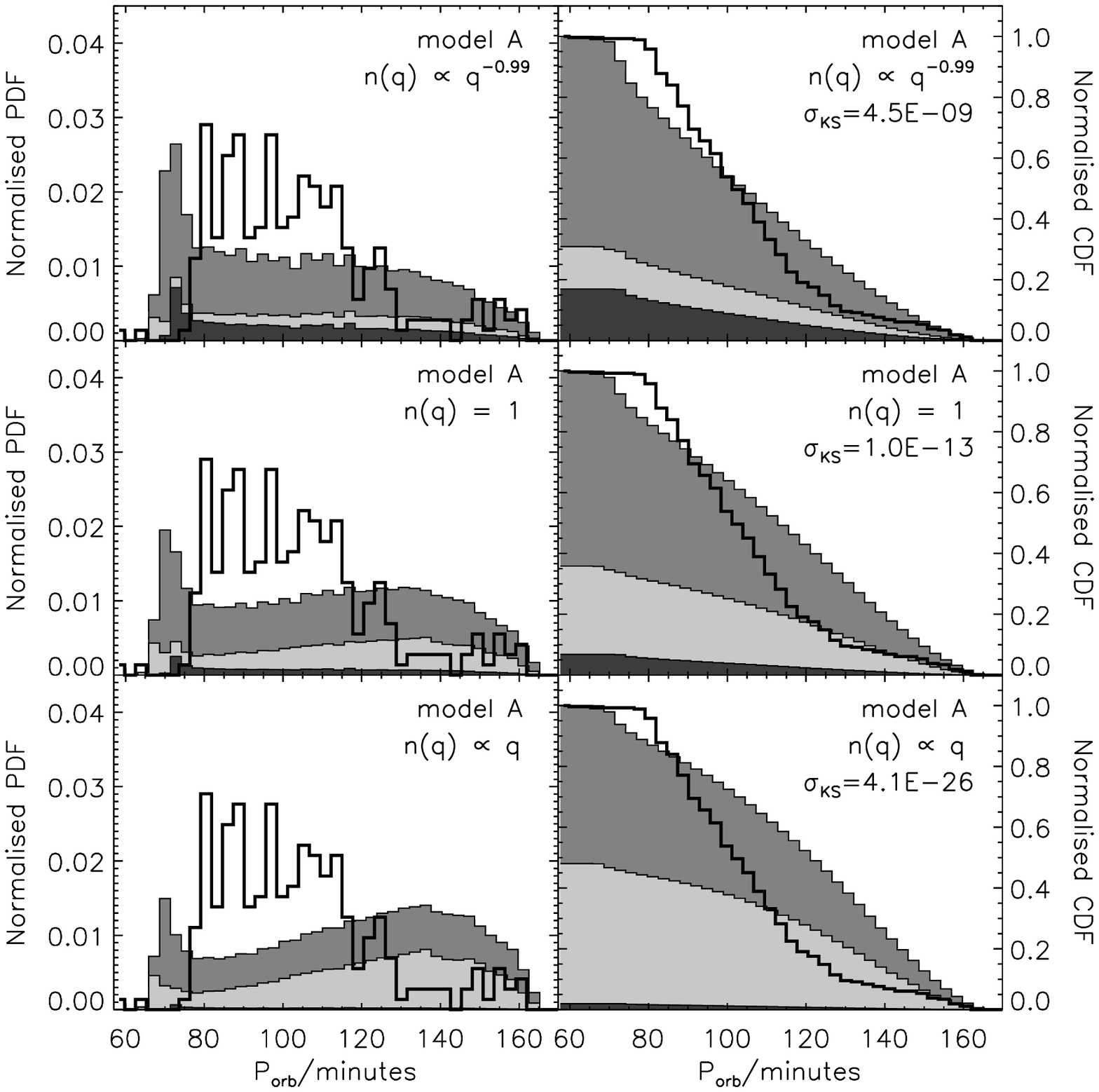}}
\resizebox{8.5cm}{!}{\includegraphics{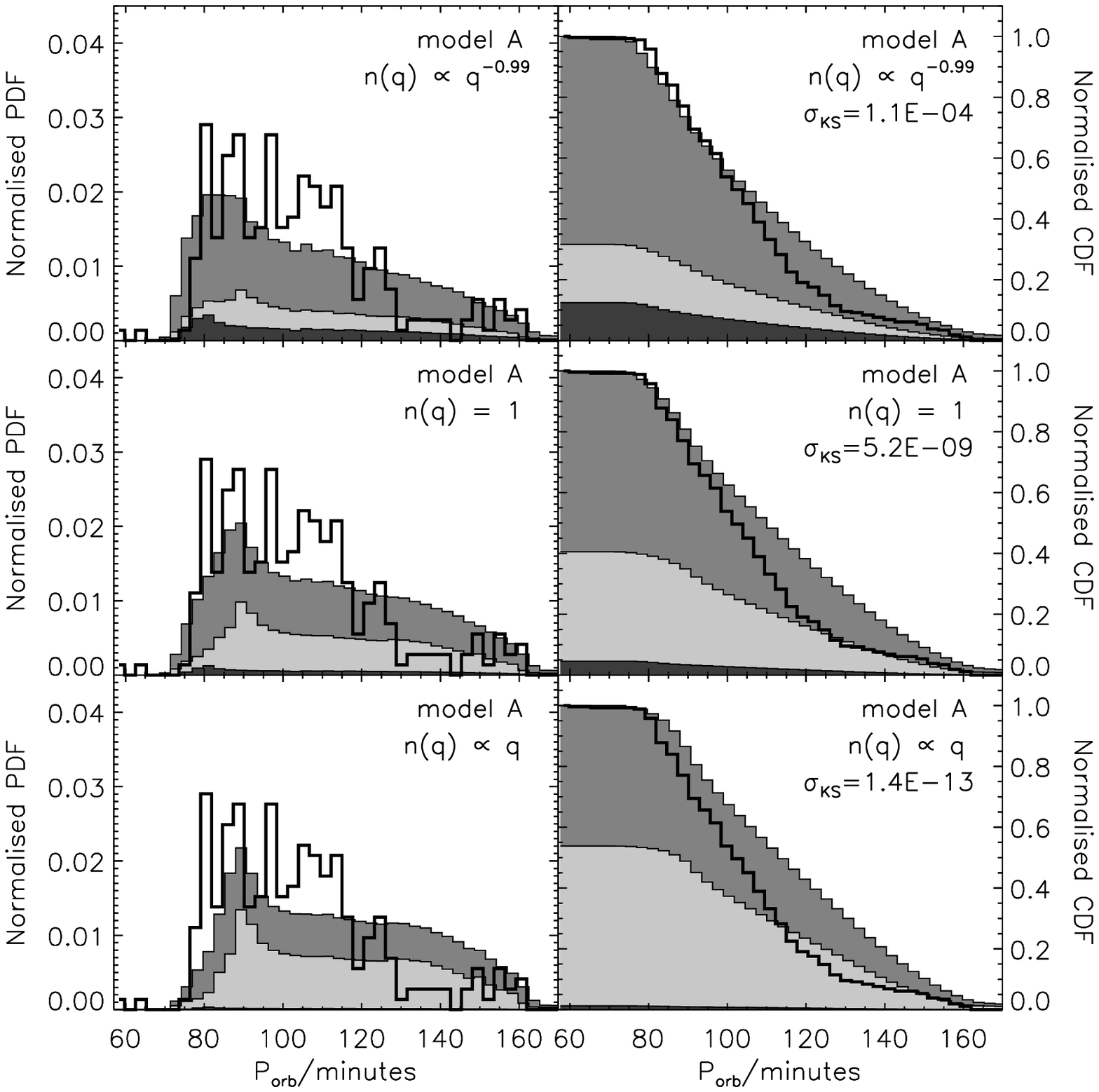}}
\caption{\small As Fig.~\ref{nodisk2}, but with the contribution of
each CV to the PDFs weighted according to the accretion luminosity to
the power 1.5. The dark gray shading indicates the contribution of
systems with O/Ne/Mg white dwarfs.}
\label{weightednodisk}
\end{figure*}

The normalized PDFs and CDFs of the present-day observed CV population
for systems without a CB disk and for population synthesis model A are
illustrated in the left most panels of Fig.~\ref{weightednodisk} for
three different initial mass ratio distributions. As discussed above,
the contribution of each system to the PDFs and the CDFs is weighted
according to the accretion luminosity to the power 1.5.

The form of our weighting factor tends to favor systems at longer
orbital periods, removing the monotonic increase in the number
of systems from long to short orbital periods that was observed in
the intrinsic distribution (cf. Fig.~\ref{nodisk2}). Consequently, the
overabundance of systems at longer periods already found in the
intrinsic CDFs is emphasized further by the introduction of the
$L_{\rm acc}^{1.5}$ weighting factor. The spike in the PDFs near
$\simeq 70$\,min and the discrepancy with the observed period minimum
also remain present for all three initial mass ratio
distributions. For $n(q)=1$, the inclusion of the weighting factor
introduces a feature in the orbital period distribution which can
either be interpreted as a dip near $\simeq 80$\,min or a hump at
$\simeq 130$\,min.  The orbital period distribution for $n(q) \propto
q^{-0.99}$ looks very similar to the distribution found for $n(q)=1$,
except for a slight increase in the relative number of systems between
$\simeq 80$ and $\simeq 100$\,min. This increase tends to make the dip
at $\simeq 80$\,min disappear and render the PDF fairly flat between
$\simeq 80$ and $\simeq 130$\,min. The most striking effect of the
initial mass ratio distribution is to enhance the contribution of
systems with more massive WDs (in addition to the enhancement due to
the weighting factor). Systems with O/Ne/Mg WDs, in particular, show
an appreciable contribution when $n(q) \propto q^{-0.99}$.
Correspondingly, the relative contribution of systems with He WDs
becomes small. We note that, as before, the assumption that $M_2$ is
distributed independently according to the same IMF as $M_1$ leads to
nearly the same distribution as for $n(q) \propto q^{-0.99}$. The
largest differences with respect to $n(q)=1$ occur for $n(q) \propto
q$. Here, the dip at $\simeq 80$\,min (or the hump at $\simeq
130$\,min) is very pronounced (in agreement with Kolb
1993). Furthermore, there are hardly any systems with O/Ne/Mg WDs, and
the relative contribution of systems with He WDs becomes noticeably
larger.

\subsubsection{CVs evolving under the influence of a CB disk}

The PDFs and CDFs for CVs with a CB disk which account for
observational selection effects are illustrated in the right most
panels of Fig.~\ref{weightednodisk}.  The main trend in the intrinsic
distributions is preserved with the weighting factor.  That is, there
is a monotonic increase in the number of systems from longer to
shorter periods, leading to the broad peak in the orbital period
distribution with maximum at $\simeq 80-90$\,min. From the CDFs it is
furthermore clear that the theoretical distributions systematically
overestimate the relative number of systems at periods longer than
$\simeq 90$\,min for all three initial mass ratio distributions. As
for the no-disk case, the initial mass ratio distribution $n(q)
\propto q^{-0.99}$ (as well as an independent $M_2$ distribution),
emphasizes the tendency of the $L_{\rm acc}^{1.5}$ weighting factor to
favor systems with heavier WDs even more and to disfavor systems with
He WDs. The initial mass ratio distribution $n(q) \propto q$ has the
opposite effect. In this case, the monotonically increasing behavior
from long to short periods is furthermore replaced by a flat plateau
ranging from $\simeq 100$ to $\simeq 130$\,min.

\subsubsection{The role of a flow of systems from above the gap}

Regardless of whether or not a CB disk is included in the simulations,
the models presented so far all fail to reproduce the steep lower edge
of the period gap near 2.25\,hrs. This may in part be caused by our
assumption that no systems formed at periods longer than 2.75\,hrs
evolve to periods below 2.75\,hrs.

To examine the effect of a flow of systems from periods longer than
2.75\,hrs (i.e., above the period gap), we multiplied the birth rate
of systems with C/O WDs born at 2.25\,hrs (135\,min) by a factor of
100 (for all C/O WD and donor masses) (cf. Kolb \& Baraffe
1999)\footnote{In practical terms, the artificial increase of the
birth rate of systems forming at 2.25\,hrs was done by multiplying the
number of systems formed in the 2.8\,min bin centered on 2.25\,hrs by a
factor of 100.}. In the case of a flat initial mass ratio distribution
$n(q)=1$, the factor roughly corresponds to the ratio of the total
birth rate of all C/O WD CVs forming at periods longer than 2.75\,hrs
to the birth rate of C/O WD CVs forming at 2.25\,hrs, and therefore
mimics a flow consisting of all systems with main sequence donors
forming above the period gap. We note, however, that this is based on
the assumption of a constant magnetic braking time scale of 10\,Gyr,
which probably underestimates the strength of magnetic braking for
systems forming above the period gap. The derived birth rates for CVs
forming at periods longer than 2.75\,hrs should therefore be
considered as lower limits, so that the simulated flow of systems from
above the gap corresponds to only a fraction of the available
long-period systems\footnote{Note that this reasoning could be
inverted to derive a lower limit on the formation rate of systems above 
2.75\,hrs by fitting the theoretical orbital period distribution as a 
function of the magnitude of the flow of systems from above the gap.
Assuming that magnetic braking and gravitational radiation are the 
only sources of orbital angular momentum losses contributing to the 
orbital shrinkage after the formation of the WD, this would place 
a lower bound on the strength of magnetic braking for systems with 
orbital periods longer than 2.75\,hrs. Such a study is, however, 
beyond the scope of this investigation.}. As it stands, in the 
case of $n(q)=1$ and population
synthesis model~A, the artificial flow furthermore implies that there
are about $\sim 1.5$ times more systems evolving from above to below
the gap than there are systems being formed below the gap.

\begin{figure*}
\rotatebox{90}{\resizebox{9.0cm}{!}{\includegraphics{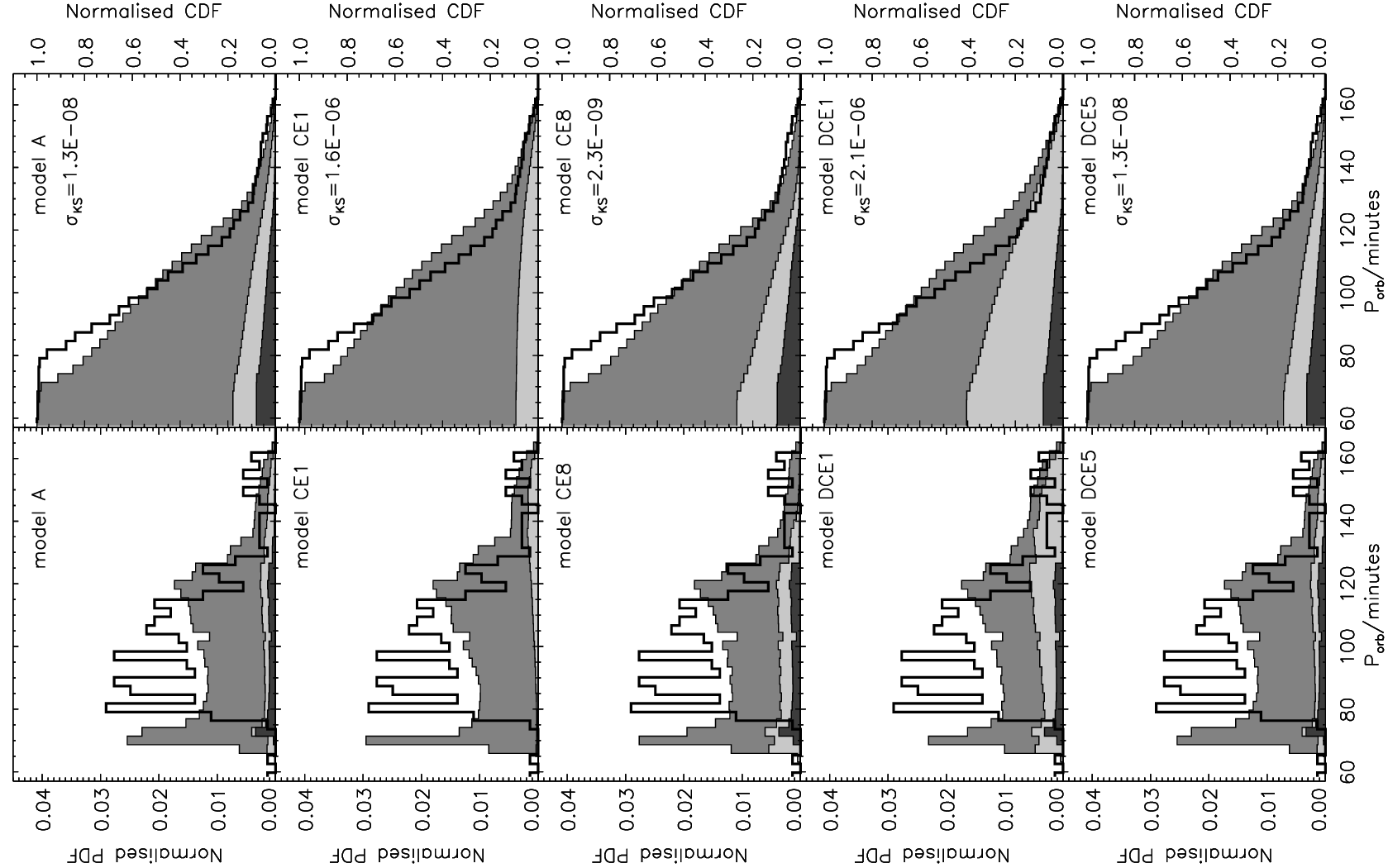}}
\resizebox{9.0cm}{!}{\includegraphics{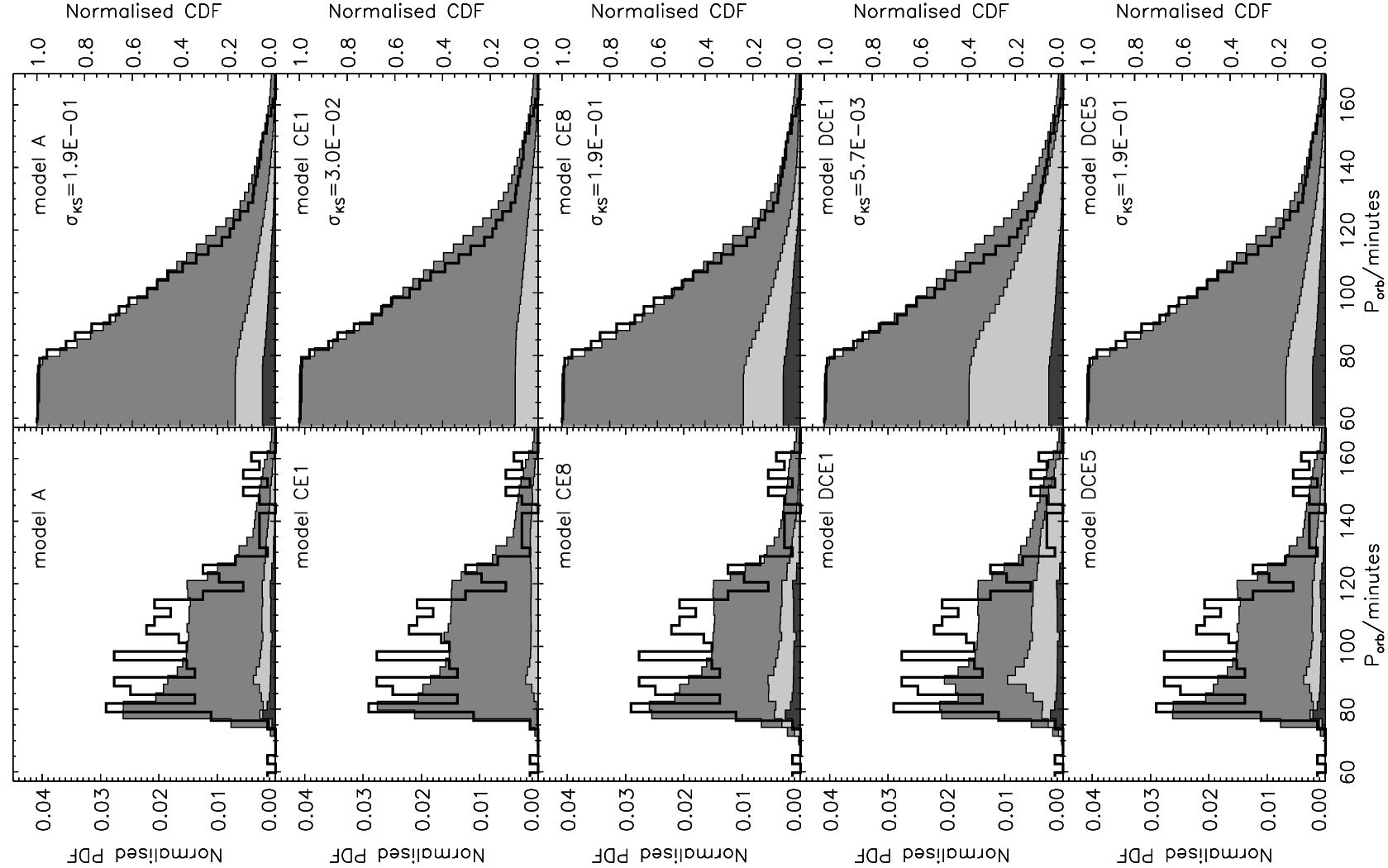}}}
\caption{\small Orbital period distribution of Galactic CVs for an
initial mass ratio distribution $n(q)=1$ and different CE models. A
flow from systems above the gap is simulated by multiplying the birth
rate of systems with C/O WDs born at 2.25h (135min) by a factor of 100
(for all C/O WD and donor masses).  The contribution of each system to
the PDFs is weighted according to the accretion luminosity to the
power 1.5.  The left (right) most panels correspond to evolutionary
tracks without (with) a CB disk for different CE models.  The light,
intermediate, and dark shading represent contributions of the He, C/O,
and O/Ne/Mg WDs to the population. The agreement between the observed and
simulated CDFs is indicated by the Kolmogorov-Smirnov significance
level $\sigma_{\rm KS}$.}
\label{weightednodiskx100}
\end{figure*}

\begin{figure*}
\resizebox{8.5cm}{!}{\includegraphics{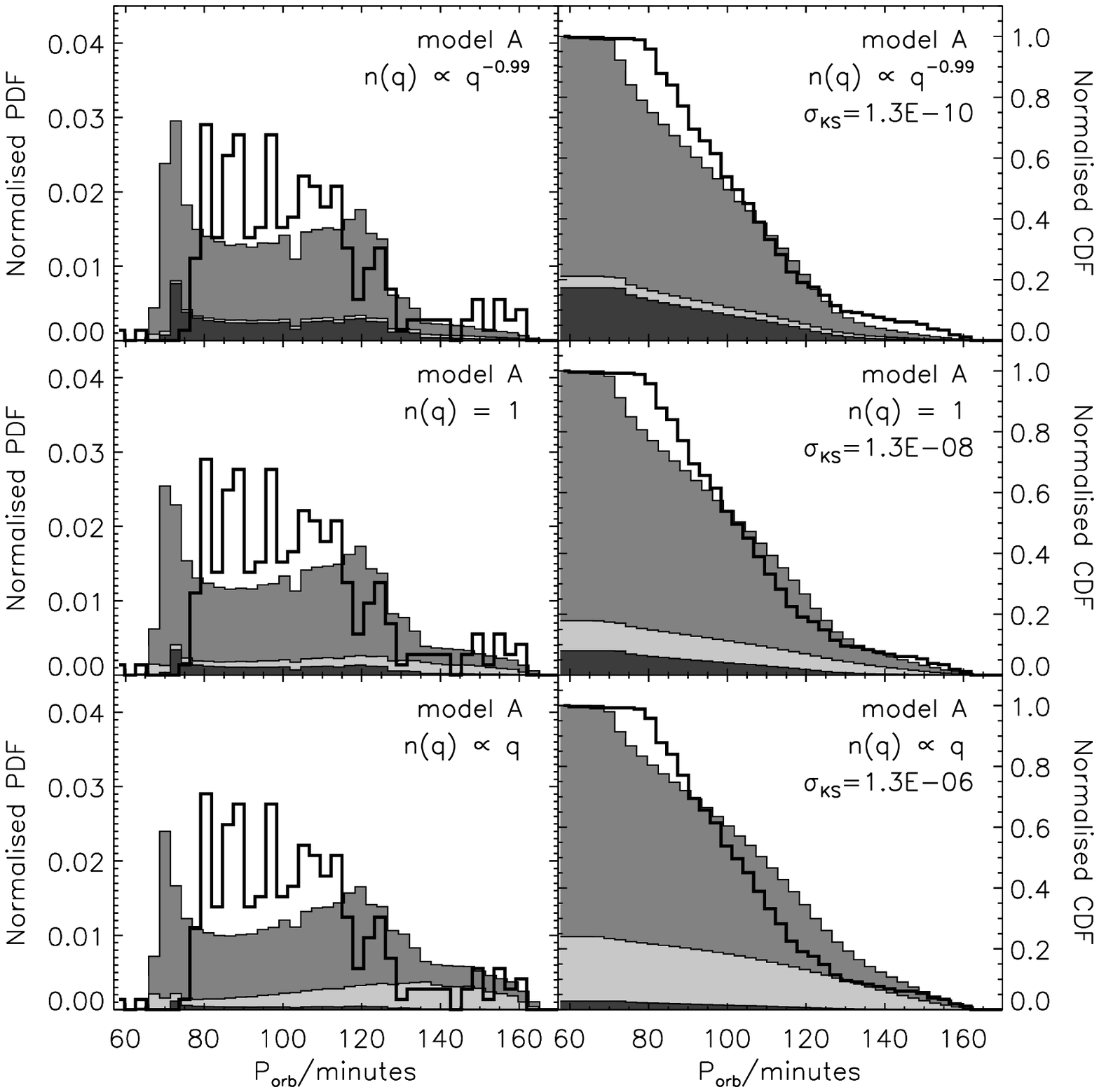}}
\resizebox{8.5cm}{!}{\includegraphics{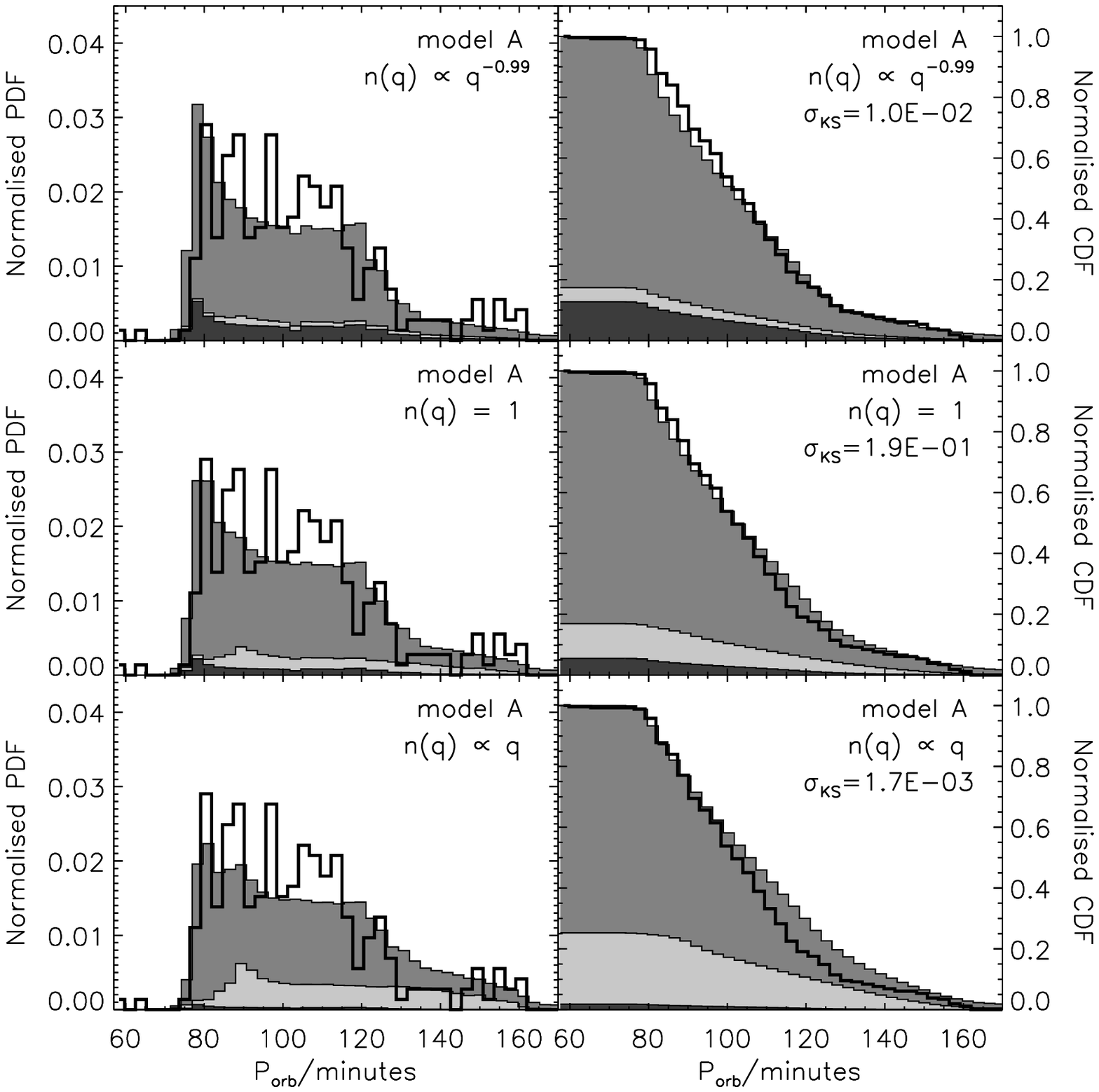}}
\caption{\small As Fig.~\ref{weightednodiskx100}, but for population synthesis
  model A and different initial mass ratio distributions.} 
\label{weightednodiskx100b}
\end{figure*}

The resulting normalized PDFs and CDFs of the present-day observed CV
population for systems without a CB disk and for the initial mass
ratio distribution $n(q)=1$ are shown in the left most panels of
Fig.~\ref{weightednodiskx100}. As above, the contribution of each
system to the distributions is weighted according to the accretion
luminosity to the power 1.5.  Compared to the PDFs without a flow of
systems from above the gap, there is a relative increase in the
number of systems at the period spike near $\simeq 70$\,min and at
$\simeq 120$\,min, with a corresponding relative decrease in the
number of systems in the period gap between 2.25 and
2.75\,hrs. Similar tendencies are observed for other initial mass
ratio distributions, as shown in Fig.~\ref{weightednodiskx100b}.  For
all models, artificially increasing the formation rate near 2.25\,hrs
results in the bump at $\simeq 120$\,min becoming more pronounced when
the number of systems assumed to form in this period bin is increased.
This effectively creates a valley at the position of the observed
period gap, improving the agreement with the observed orbital period
distribution. The improvement is particularly striking for models A,
CE8, and DCE5. In the case of population synthesis model A, adopting
$n(q) \propto q^{-0.99}$ (or distributing $M_2$ independently
according to the same IMF as $M_1)$, enhances the similarity between
theoretical and observed distributions even further, resulting in a
good representation of the lower edge of the period gap. A similar
conclusion applies to the other population synthesis models. In the
case of the initial mass ratio distribution $n(q) \propto q$,
artificially increasing the formation rate near 2.25\,hrs by a factor
of 100 is not enough to create a sharp increase in the number of
systems at the lower edge of the period gap.

The PDFs and CDFs of the orbital periods of CVs with a CB disk,
obtained by artificially increasing the formation rate near 2.25\,hrs
by a factor of 100 and by adopting the initial mass ratio distribution
$n(q) = 1$, are shown in the right most panels of
Fig.~\ref{weightednodiskx100}. The effect of different initial mass
ratio distributions is illustrated in the right most panels of
Fig.~\ref{weightednodiskx100b}. The flow of systems at 2.25\,hrs has
two main effects. As for the no-disk case there is an increase in
the number of systems at $\simeq 120$\,min which significantly
contributes to the creation of a gap in the PDFs at periods above
$\simeq 120$\,min. Here, the increase does not lead to a local peak in
the distribution functions though, but instead further flattens the
PDFs between $\simeq 90$ and $\simeq 120$\,min. In addition, the rise
of the short-period peak from $\simeq 70$ to $\simeq 80-90$\,min tends
to become steeper than when no flow of systems at 2.25\,hrs is
considered (cf. Fig.~\ref{weightednodisk}). This is mainly due to the
period bounce of the evolutionary tracks with initial periods near
2.25\,hrs. The significance level of the Kolmogorov-Smirnov statistic
furthermore indicates that the overall agreement with the observed
PDFs and CDFs is best for population synthesis models A, CE8, and
DCE5, and for the flat initial mass ratio distribution $n(q) = 1$,
although the lower edge of the period gap seems to be modeled best by
$n(q) \propto q^{-0.99}$. It is interesting to note that spreading out
the artificial flow of systems over several period bins near 2.25\,hrs
does not significantly affect any of these results. In particular, a
test-run in which the artificial flow of systems was spread out over a
range of $\simeq 14$\,min centered on the period of 2.25\,hrs, yielded
PDFs and CDFs that are almost indistinguishable from the ones obtained
by concentrating the entire flow to the 2.25\,hrs period bin.

\subsubsection{Summary}

In order to examine the effect of observational biases in the observed 
population of Galactic CVs, the contribution of simulated systems to the 
theoretical orbital period distributions was weighted according to the 
accretion luminosity raised to the power 1.5. Such a weighting leads 
to an increased contribution of systems at orbital periods immediately 
below the lower edge of the period gap.  The distribution of systems 
near the period minimum, however, is not affected perceptively since 
the shift of the period minimum and the smearing out of the number of 
systems near the period minimum with the additional angular momentum loss 
is preserved when the observational selection effects are included.  
On the other hand, the simulations fail to reproduce the steep lower edge of 
the gap in the observed orbital period distribution. The agreement can be 
improved by increasing the number of systems "formed" at periods of 
$\simeq 2.25$\, hrs, possibly indicating that a significant fraction of 
systems formed at periods longer than 2.75\,hrs contribute to the short 
period population.

\subsection{The distribution of mass-transfer rates and donor masses as a 
function of orbital period}

\begin{figure*}
\resizebox{8.3cm}{!}{\includegraphics{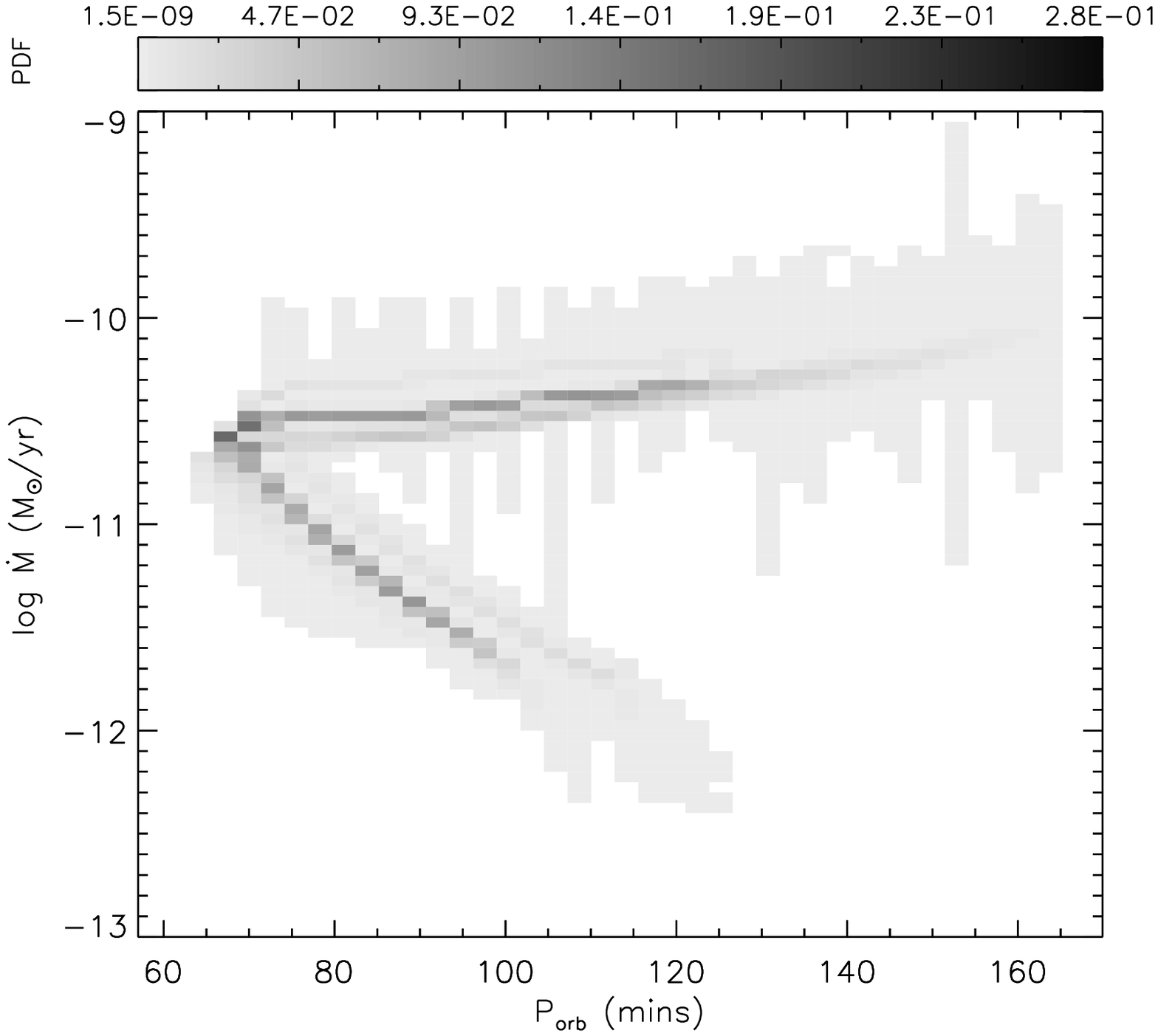}}
\resizebox{8.3cm}{!}{\includegraphics{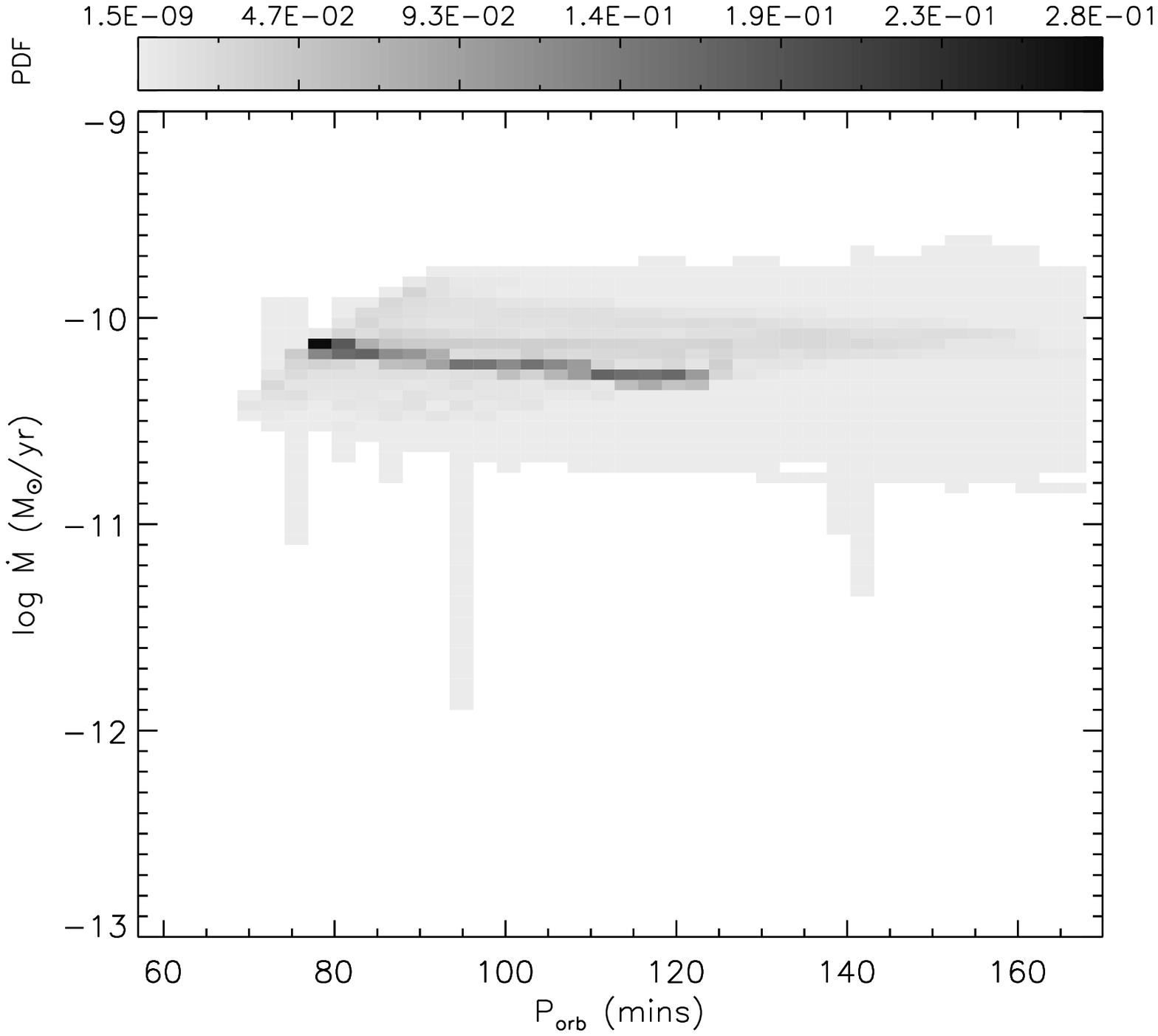}}
\caption{Normalized PDFs of the intrinsic present-day CV population in
the $(\dot{M_{\rm d}},P_{\rm orb})$-plane, for population synthesis
model~A and the initial mass ratio distribution $n(q)=1$.  A flow from
systems above the gap is simulated by multiplying the birth rate of
systems with C/O WDs born at 2.25h (135min) by a factor of 100. The
left (right) most panels correspond to evolutionary tracks without
(with) a CB disk.}
\label{mdot1}
\end{figure*}

In Fig.~\ref{mdot1}, we show the intrinsic distribution of
mass-transfer rates as a function of orbital period, for population
synthesis model~A and for a flat initial mass ratio distribution
$n(q)=1$. A flow of systems forming above the gap is simulated as
above by multiplying the birth rate of systems with C/O WDs born at
2.25h by a factor of 100. In the case of CVs without a CB disk, the
evolution under the influence of gravitational radiation clearly
converges to two distinct tracks, with the dominant one 
associated with the systems containing C/O WDs (cf. Howell, Nelson, \&
Rappaport 2001). Mass-transfer rates are typically of the order of
$3-5 \times 10^{-11}\,M_\odot\,{\rm yr^{-1}}$ for systems evolving
from longer to shorter periods. The rates decrease rapidly with
increasing orbital period for systems that have evolved beyond the
minimum period, reaching values of the order of $3 \times
10^{-12}\,M_\odot\,{\rm yr^{-1}}$ at periods of $\simeq 100$\,min. In
the case of CVs with a CB disk, the typical mass-transfer rates are of
the order of $5-8 \times 10^{-11}\,M_\odot\,{\rm yr^{-1}}$. As a direct 
observational consequence, the post-bounce systems evolving away from the 
period minimum due to the combined effect of gravitational radiation and 
a CB disk can have mass transfer rates as much as an order of magnitude 
higher than that promoted by gravitational radiation only. 

\begin{figure*}
\resizebox{8.3cm}{!}{\includegraphics{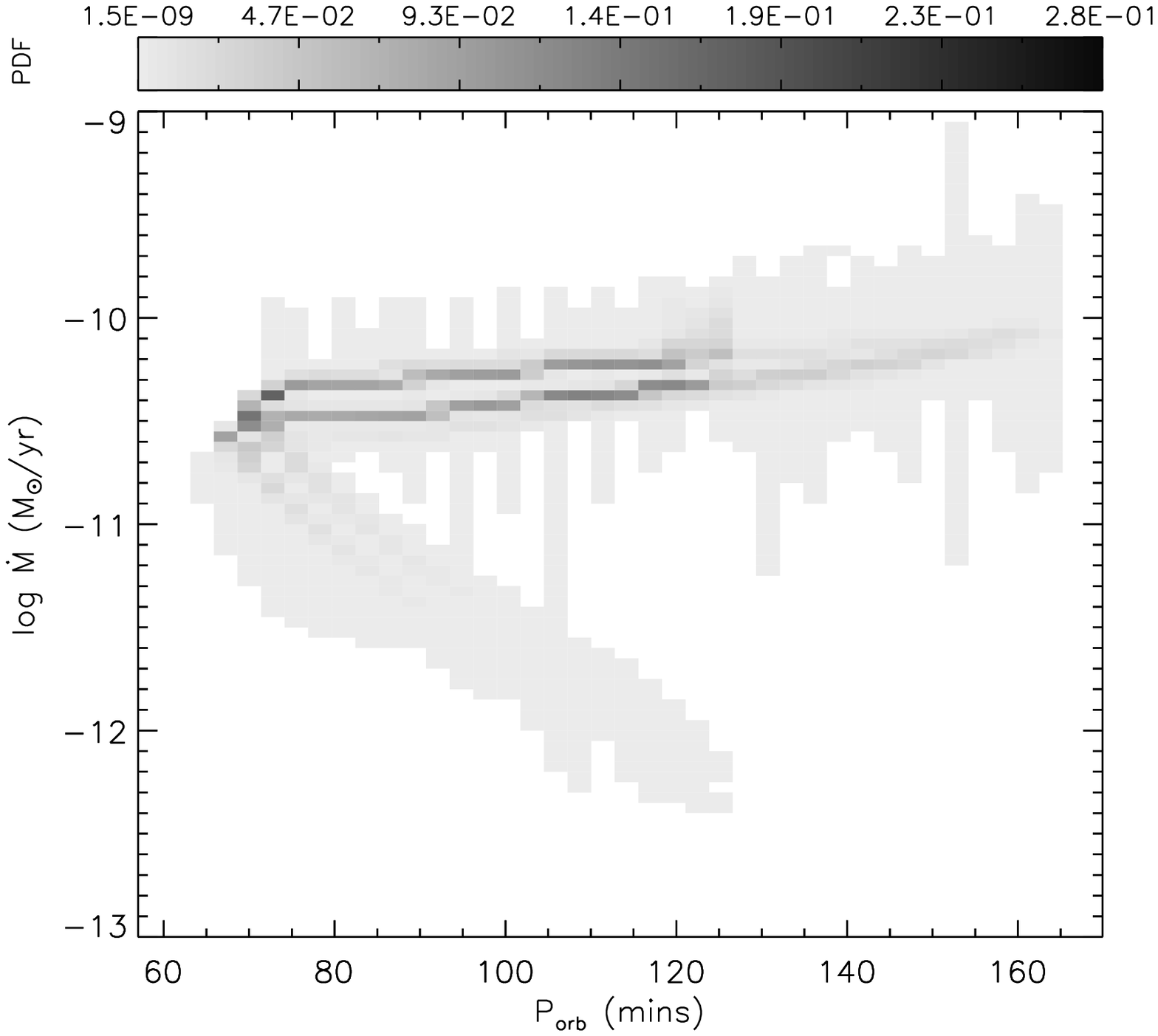}}
\resizebox{8.3cm}{!}{\includegraphics{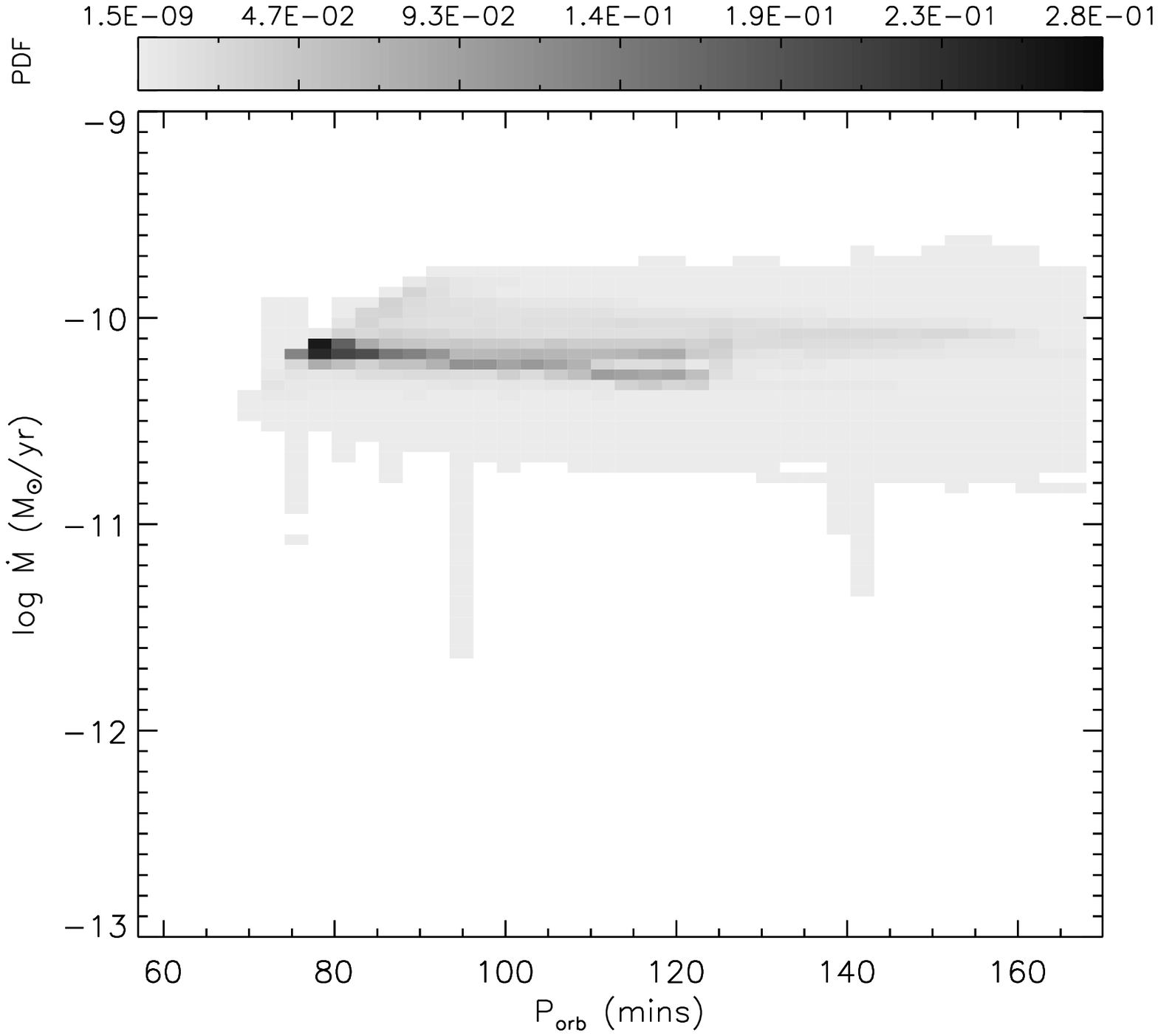}}
\caption{As Fig.~\ref{mdot1}, but for the present-day CV population
obtained by weighting the contribution of each system to the PDF
according to $L_{\rm acc}^{1.5}$. }
\label{mdot2}
\end{figure*}

In contrast to the population without a CB disk, the mass-transfer
rates increase slightly with decreasing orbital periods.  The
higher rates are, in fact, responsible for the shift in the
theoretical orbital period minimum to longer periods in comparison to
models with gravitational wave angular momentum losses only and are
caused by the dependency of the mass-transfer rate on the mass of the
CB disk (see Fig. 6 of Taam et al. 2003). For a given orbital period,
the mass-transfer rates are furthermore spread over a wider range of
values than in the case of CVs evolving under the influence of
gravitational radiation only.  Weighting the contribution of the
systems to the PDFs according to the accretion luminosity to the power
1.5, as is done in Fig.~\ref{mdot2}, favors systems at longer orbital
periods as a result of higher mass transfer rates (for the population
evolving without a CB disk) and higher typical WD masses at longer
periods, reflecting the differences between the He WD and C/O WD
tracks (for both populations). In addition, the weighting factor has a
consequence of decreasing the contribution of post-bounce systems for
a CV population evolving without a CB disk.

\begin{figure*}
\resizebox{8.3cm}{!}{\includegraphics{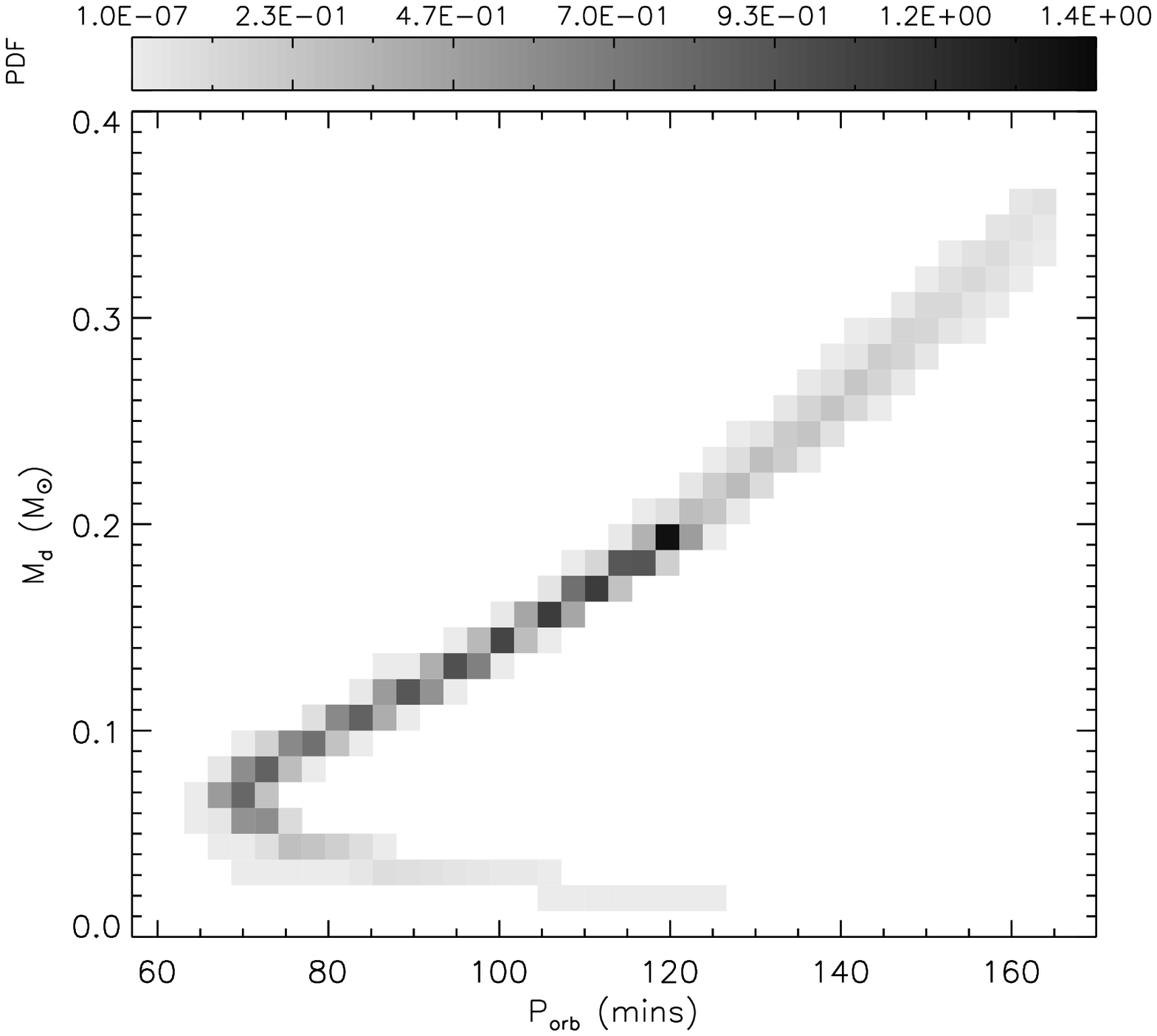}}
\resizebox{8.3cm}{!}{\includegraphics{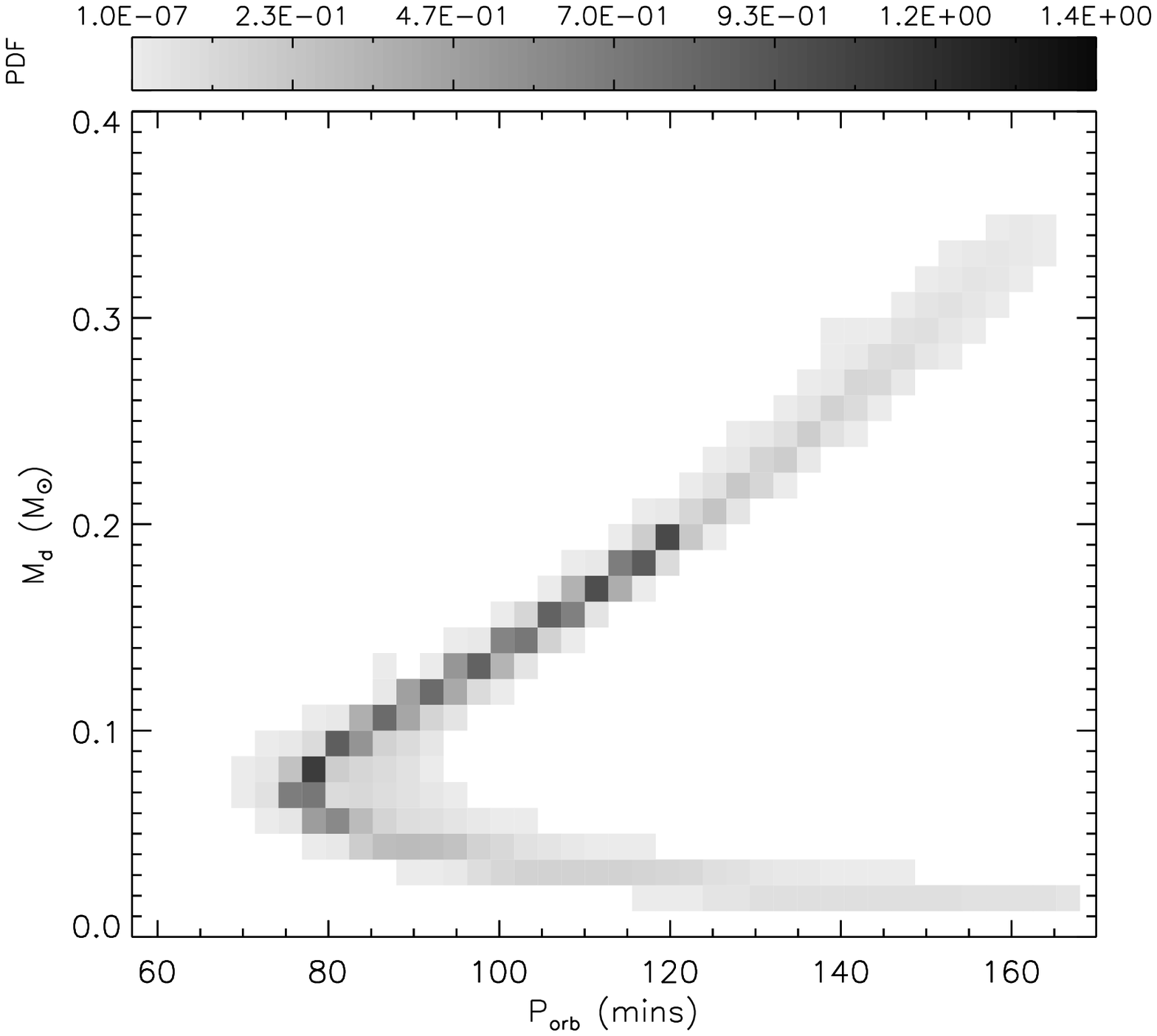}}
\caption{Normalized PDFs of the present-day CV population in
the $(M_{\rm d},P_{\rm orb})$-plane, for population synthesis
model~A and the initial mass ratio distribution $n(q)=1$.  A flow from
systems above the gap is simulated by multiplying the birth rate of
systems with C/O WDs born at 2.25h (135min) by a factor of 100. The
left (right) most panels correspond to evolutionary tracks without
(with) a CB disk.  The present-day CV population is
obtained by weighting the contribution of each system to the PDF
according to $L_{\rm acc}^{1.5}$.}
\label{mdon2}
\end{figure*}

The distribution of donor masses is displayed as a function of orbital
period in Fig.~\ref{mdon2} for the simulated distributions obtained by
incorporating an artificial flow of systems from above the gap
and by weighting the contribution of each system according to its
accretion luminosity raised to the power 1.5. The larger spread of the
systems, both in $M_{\rm d}$ and $P_{\rm orb}$, when the effects of a
CB disk are taken into account again illustrates that the evolutionary
sequences in this case do not converge to a common track. Typical
donor star masses near the minimum period are similar for the two
populations ranging from 0.05 to 0.09\,$M_\odot$. In the case of CVs
with a CB disk, the post-bounce systems can furthermore evolve to
significantly longer orbital periods than systems without a CB disk.

Hence, our simulations predict that the presence of a CB disk
gives rise to a population of CVs with low-mass donor stars at
significantly longer orbital periods and higher mass-transfer rate
than is possible without a CB disk. It is to be noted though that this
population constitutes a low-density tail in the $\left( M_d, P_{\rm
orb} \right)$ parameter space. Statistically, observational
confirmation of the existence of these systems can therefore be
expected to be quite challenging.

\subsection{Typical CB disk properties}

\begin{figure*}
\resizebox{8.3cm}{!}{\includegraphics{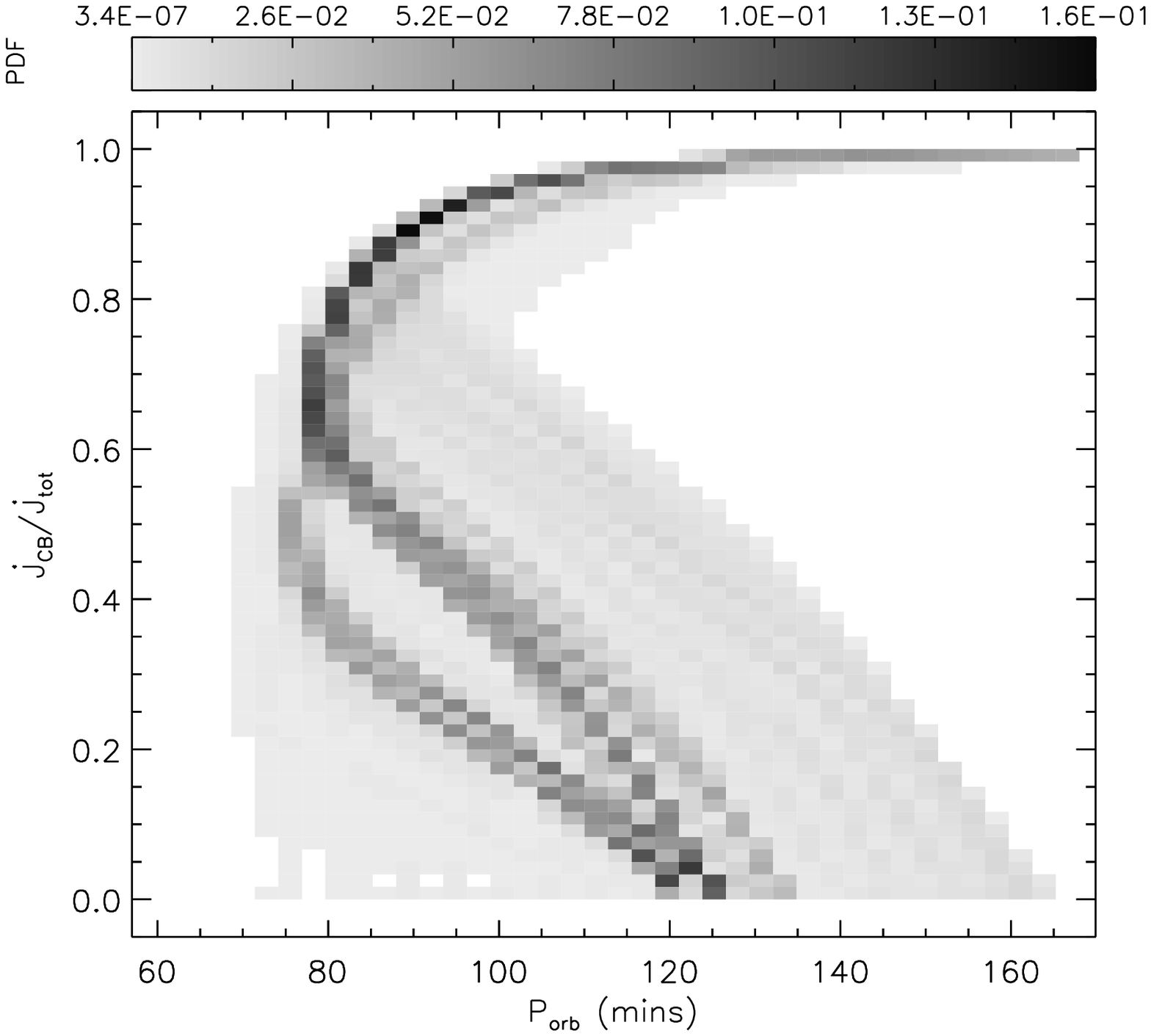}}
\resizebox{8.3cm}{!}{\includegraphics{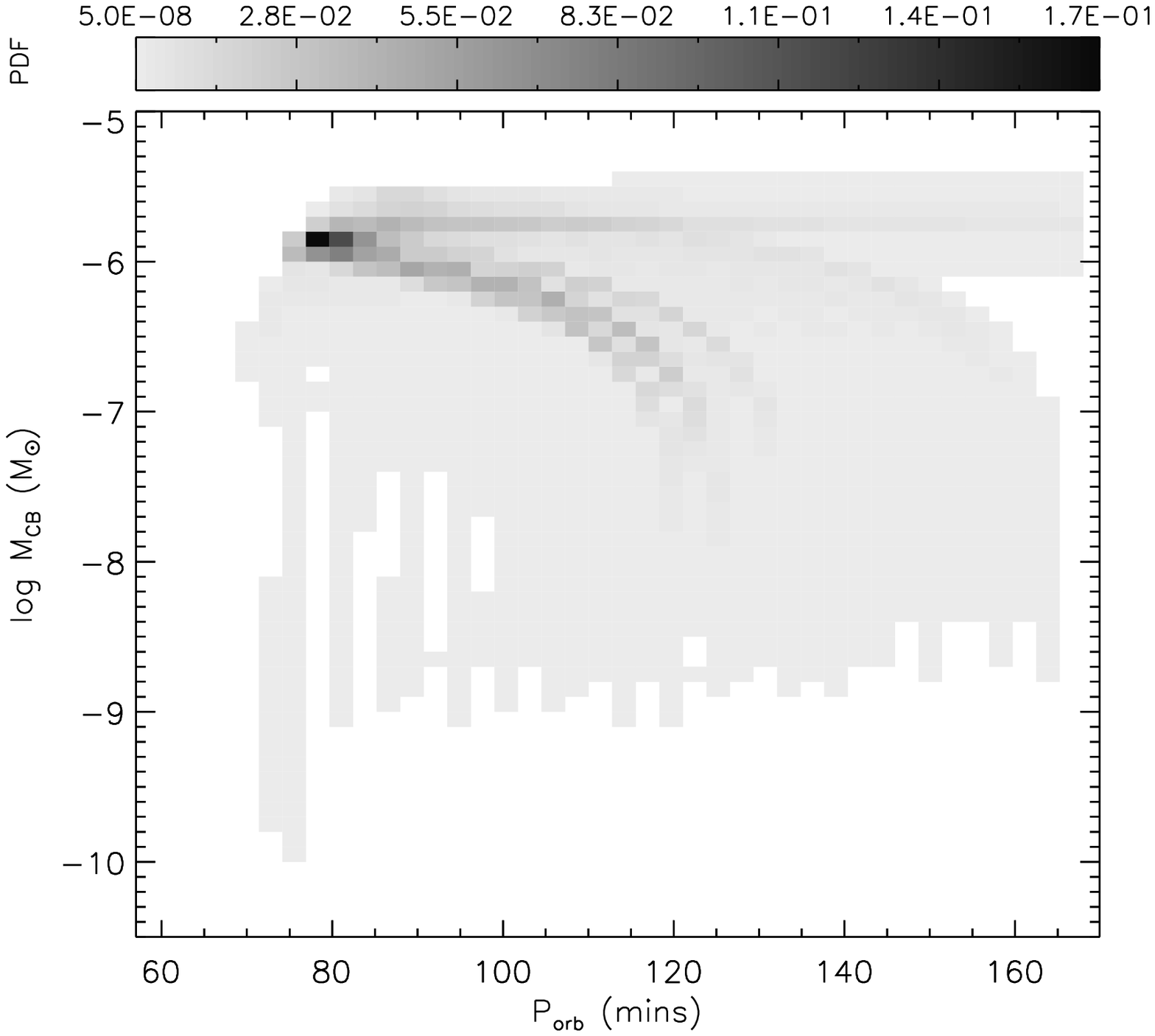}}
\caption{Normalized PDFs of the present-day CV population in the
$(\dot{J}_{\rm CB}/\dot{J}_{\rm tot}, P_{\rm orb})$- and $(M_{\rm CB},
P_{\rm orb})$-planes, for population synthesis model~A and the initial
mass ratio distribution $n(q)=1$ in the left and right most panels 
respectively.  A flow from systems above the gap
is simulated by multiplying the birth rate of systems with C/O WDs
born at 2.25h (135min) by a factor of 100. 
The present-day CV population is obtained by weighting the
contribution of each system to the PDF according to $L_{\rm
acc}^{1.5}$.}
\label{mcb}
\end{figure*}

The binary population synthesis allows us to investigate the typical
properties of putative CB disks in the present-day CV population. For
this purpose, we use population synthesis model~A with the initial
mass ratio distribution $n(q)=1$, including angular momentum losses to
the CB disk $\dot{J}_{\rm CB}$ and a flow of systems from above the
gap (the model illustrated in the right panels of
Figs.~\ref{mdot1}-\ref{mdon2}). In the left panel of Fig.~\ref{mcb} we
show the fraction of the total angular momentum loss caused by the CB
disk $\dot{J}_{\rm CB}/\dot{J}_{\rm tot}$ as a function of orbital
period, weighted by $L^{1.5}_{\rm acc}$. The CB disk dominates the
evolution close to and beyond the period bounce. A typical system with
$P_{\rm orb} \la 90$\,min will typically lose more than half its
angular momentum to the CB disk.  The figure also shows two branches
of pre-bounce systems with $\dot{J}_{\rm CB}/\dot{J}_{\rm tot} \la
0.5$ which are dominated by systems flowing in from above the gap. The
upper branch is already present in the intrinsic unweighted population
and corresponds to CVs with $M_{\rm WD} \approx
0.6$--$0.8\,M_\odot$. The lower branch on the other hand is associated
with $M_{\rm WD} \approx 1.2$--$1.4\,M_\odot$ and results
predominantly from the $L^{1.5}_{\rm acc}$ weighting factor favoring
systems with more massive white dwarfs.

In the right panel of Fig.~\ref{mcb}, we plot the CB disk mass $M_{\rm
CB}$ as a function of orbital period. The mass saturates at $M_{\rm
CB}\approx 10^{-6}M_\odot$ corresponding to the transfer onto the WD
of about 0.1~$M_\odot$ of material from the low mass companion
($\delta=\dot{M}_{\rm CB}/\dot{M_2}=10^{-5}$, see
\S\ref{evtracks}). Systems closest to the period minimum have the most
massive CB disks. The disks for those systems have black body
temperatures at their inner radius $\approx 1,500$\,K with surface
densities of $\approx 100\, {\rm g\,cm^{-2}}$, and a size of $\approx
0.3$~AU (defined as the radius at which the disk becomes optically
thin). For pre-bounce systems with longer periods (younger disks) the
typical temperature decreases to $\approx 800$~K at $P_{\rm
orb}=120$~min with an increasing spread in values.

\section{Conclusions}

The orbital period distribution of CVs has been investigated via a
binary population synthesis technique for a range of binary input
parameters and angular momentum loss processes.  It has been 
argued that processes such as residual magnetic braking and 
consequential angular momentum loss are unlikely to be responsible 
for the location and lack of accumulation of systems at the observed 
period minimum of 80 minutes.  As a result, we have explored  
the effect of CB disks on the evolution of systems to determine 
the generic properties of an alternative angular momentum loss 
prescription on the CV orbital period distribution.  In particular, 
we have focused on the evolution of systems below 2.25 hours to 
discriminate our study from studies concerned with the formation of 
the period gap. 

To provide input to our population synthesis study, the zero age 
CV population has been calculated for a wide range of models. A 
common feature of these models is the formation of a gap with 
systems characterized by C/O white dwarfs above the gap and a 
significant contribution from systems with He white dwarfs below 
the gap.  We note that Webbink (1979) has suggested that the origin of
the period gap of the observed CV population may be related to this
different population of white dwarfs.  However, taking into account
the evolution of CVs leads to a modification of the zero age
period distribution. 

In the population synthesis models carried out in this investigation,
the additional angular momentum loss associated with the CB disks
shift the period minimum to longer orbital periods than for models
with angular momentum losses due to gravitational radiation and 
the long-term mean mass-loss rate from the system due to nova
explosions (cf., Kolb 1993; Howell et al. 2001; Barker \& Kolb 2003).
Specifically, the minimum period shifts to the observed value at about
$\simeq 80$\,min with the number of systems sharply increasing to a
period of about $\simeq 90$\,min when account is taken of systems
forming in the period gap. On the other hand, the peak in the
distribution functions lies at $\simeq 80-85$\,min when the
contribution of systems formed in the gap is not taken into
account. This result suggests that the systems formed within the
period gap are important in determining the distribution at orbital
periods ranging from $\simeq 80-90$\,min.  We point out that there is
no evidence for an accumulation of systems at the minimum period.
This is a direct result of the fact that the evolutionary sequences,
starting from different initial orbital periods, do not converge to a
common track. In particular, the dependence of the mass-transfer rate
on the mass of the CB disks introduces a broader range of
mass-transfer rates at a given orbital period than evolutions based on
gravitational radiation and the long-term mean mass-loss rate
from the system due to nova explosions. Thus, the angular momentum
loss process associated with our alternative prescription provides an
additional ingredient allowing systems with longer initial orbital
periods (i.e., more massive donors) to undergo a period bounce at
longer periods.  This reflects the fact that the mass in the CB disk
is a function of the initial parameters of the binary system,
resulting in a slight increase (rather than a decrease) of the mass
transfer rate with decreasing orbital period.
Alternatively, this trend could also be achieved for a range of
constant mass input rates in the CB disk at a given initial orbital
period with those systems characterized by higher mass input rates
having longer bounce periods. For example, fractional mass input rates
into the CB disk as high as $10^{-4}$ would lead to bounce periods
near the orbital period at which mass transfer is initiated (see Taam
et al. 2003), whereas mass input rates significantly less than the
values adopted here would correspond to evolutions similar to those
for which gravitational radiation losses were the only angular
momentum loss mechanism considered. Based on this behavior, we 
tentatively constrain the fractional mass input rates into the CB 
disk to be in the range from $10^{-7}$ to $10^{-4}$. A more stringent 
range would require a population synthesis study of CVs with CB disk 
for a range of mass input rates. We also note that the calculated
mass-transfer rates found for the short period non magnetic CVs lie in
the range for which the accretion disk is thermally unstable (e.g.
Osaki 1996; Hameury et al. 1998).  This suggests that the permanent
superhump systems, which are believed to be systems accreting at
sufficiently high mass transfer rates such that the accretion disks
are thermally stable, may form only a very small fraction of the CV
population corresponding to higher angular momentum loss rates than
those found here or to a small fraction of the CV phase if
mass-transfer cycles occur about the secular mean values of the
mass-transfer rate (e.g., King et al. 1996; B\"uning \& Ritter 2004).

Since there do not exist complete CV surveys to a given distance,
observational selection effects should be accounted for in the
discovery of CVs.  In this study we have applied a simple
approximation for the weighting factor, in which the intrinsic period
distributions were weighted by the accretion luminosity to the power
of 1.5.  This would be appropriate for the systems below the period
gap where it is not likely that they would be discovered beyond the
thickness of the Galactic disk due to their low mass-transfer rates.

For the population of systems formed below 2.75 hours, the period
distribution immediately below the lower edge of the period gap, on
the other hand, is not in agreement with observations because the
number of systems increases gradually with decreasing orbital period.
The steepness of the observed orbital period distribution at the lower
edge of the period gap can be affected by a significant modification
of the ZACV population and evolution near the lower edge of the period
gap, or by flow of systems from orbital periods above the period
gap. To avoid fine tuning in the parameters governing the ZACV
population and angular momentum loss process, we have considered the
possibility of a flow of systems from longer orbital periods.

Accordingly, we have explored the possibility of increasing the
formation rate at 2.25 hours so that the number of systems from the
lower edge of the period gap to shorter orbital periods can be
increased more sharply.  We have found that the inclusion of this flow
does not significantly affect the position of the peak, shifting it
from $\simeq 90$\,min to about $\simeq 80-85$\,min; however, the
decrease in the number of systems to the minimum period is steeper and
is in better agreement with observation. In addition, the inclusion of
the flow from orbital periods above the period gap can result in a
sharp increase in the number of systems from 2.25\,hrs to 2\,hrs. In
this case, the overall properties of the period distribution obtained
from the population synthesis models are remarkably similar to that
observed, suggesting that evolution with additional angular momentum
losses coupled with a flow from above the gap is necessary in the
lowest order of approximation.  We point out that although our results
give an indication of the effect of a flow of CVs from longer orbital
periods, they do not constrain the fraction of CVs which evolve from
above the period gap to below the period gap since it depends on the
assumptions of the orbital braking mechanism.  The evolution of CVs
above the gap with both main sequence and evolved donors driven by
magnetic braking processes operating on timescales shorter than
$10^{10}$ years would need to be included in our population synthesis
study to address this issue.

Although the qualitative description of the CV period distributions
below the period gap is not particularly sensitive to the mass ratio
distribution of the progenitor binary population and to the population
synthesis model parameters, the quantitative description is sensitive
to these inputs. In order to constrain the particular form of the
distribution and binary evolutionary parameters, a treatment of the CV
evolution above the period gap coupled with CV evolution below the gap
as explored in this paper will be necessary.  In this context,
additional constraints on the evolution are provided by the existence
of the period gap and the higher mass-transfer rates for systems above
the gap in comparison to those systems below the gap (e.g., Patterson
1984; McDermott \& Taam 1989; Kolb \& Stehle 1996; Kolb, King, \&
Ritter 1998; Howell, Nelson, \& Rappaport 2001; Townsley \& Bildsten
2005).

Other comparisons of the results of our study with the observed
properties of the short period CVs are possible, however the inherent 
faintness of the population makes such comparisons difficult to quantify.  
Of the observed properties, donor masses have been inferred for systems 
in which superhump periods have been observed. For example, 
Patterson (1998) has inferred
substellar mass donors in the range of $0.02 - 0.04 M_{\odot}$ for 4
systems with orbital periods near the period minimum.  This is
somewhat lower than the masses obtained in our calculations which
exceed $0.05\,M_{\odot}$ near the period minimum.  We note that this
minimum mass is independent of the angular momentum loss treatment,
although the period minimum differs between the two simulations. For a
given donor star mass, systems that have evolved beyond the minimum
period are furthermore characterized by a wider range of orbital
periods in the case where a CB disk is present than in the case where
the systemic angular momentum loss is governed by gravitational
radiation and the long-term mean mass-loss rate from the system due to
nova explosions only.

Since the shift in the minimum period is directly affected by the
angular momentum loss process, the higher mass-transfer rates promoted
for the pre-and post-bounce systems are a specific prediction of our
study.  The greatest deviation between models with our additional
angular momentum loss prescription from those with gravitational
radiation and the effect of the long-term mean mass loss from the
system in nova explosions are for systems near the period minimum and
during the post bounce phase of evolution.  The deviation in the
mass-transfer rates can amount to a factor of 2 near the period
minimum to an order of magnitude for post bounce systems lying near
100 minutes.  The donors in these systems lose mass at rates greater
than about $4 \times 10^{-11}\, {\rm M_{\odot} yr^{-1}}$. For 
the observed systems with rates below these values, the difference between 
the instantaneous mass transfer rates and the secular rates must be 
considered. The secular mass transfer rates determined in our 
evolutionary computations are generally difficult to measure in CVs,
especially if they undergo dwarf nova outbursts.  However, the white
dwarf can be used as a diagnostic to infer the time averaged rate of
mass accretion (see Townsley \& Bildsten 2003).  In this case,
compressional heating contributes to the thermal energy balance in the
white dwarf envelope and the white dwarf accretors in the post bounce
systems are expected to be hotter in comparison to systems evolving
under the action of gravitational radiation alone.

The results from our population synthesis suggest that the best
strategy for discovering a CB disk below the period gap would be to
target systems close to the period minimum. These will have the
highest CB disk mass ($\approx 10^{-6}$\,M$_\odot$), temperatures
($\approx 1,500$\,K), and radii ($\approx 0.3$\,AU). Dubus et
al. (2004) searched for thermal CB emission in the IR for two CVs
below the period gap, WZ~Sge ($P_{\rm orb}=76$\,minutes) and HU~Aqr
($P_{\rm orb}=125$\,minutes). For a steady-state like CB disk density
profile they put upper limits on the inner disk temperature of 800~K
and 1,100~K respectively. The fact that these temperatures are
lower than our predicted $\approx 1,500$\,K may suggest a radial
temperature profile in which the disk departs from a steady state
description. Searching for narrow absorption lines may be more suited
to finding cold CB material.  Belle et al. (2004) searched for such
lines in EX~Hya ($P_{\rm orb}=98$\,minutes) and derived an upper limit
on the column density of $10^{17}$~cm$^{-1}$. Although the inclination
of the system is high ($i\approx78^o$) it may still be insufficient to
significantly probe the cold CB material, which is expected to have a
very narrow and non constant flare angle. Of those three systems
only WZ Sge has a period in the range where an evolutionary important
CB disk could (statistically) be expected. More observations of
systems close to the period minimum would be useful.

Assuming that the additional angular momentum process is related to the 
existence of a CB disk, the source of its mass is likely to come
directly from the donor star.  Although magnetically driven mass
outflows at velocities less than the escape speed are possible from
the outer regions of the accretion disk surrounding the WD (Proga
2003), the observational fact that magnetic CVs have a similar orbital
period distribution to the non magnetic CVs suggests that such
outflows are not important for these systems since polars do not have
inner disks. On the other hand, equatorial mass loss from the 
donor may be enhanced by the mass in the region trapped by magnetic flux 
tubes that form closed loops in the magnetosphere (so called dead 
zones; Mestel 1968).  This matter may be accelerated by the centrifugal 
effect at the high rotation rates characteristic of short period systems.  
Since the dead zones of single stars rotating rapidly may be situated 
outside the outer Lagrangian points of binary systems, one might  
plausibly anticipate additional mass loss in the equatorial plane, 
possibly feeding the CB disk.

The amount of mass lost by the donor to such a disk may significantly
exceed the overall mass in the disk ($\sim 10^{-6} M_{\odot}$) if nova
outbursts take place in the system.  For the low mass-transfer rates
characteristic of the systems below the period gap, it is likely that
nova outbursts take place, but with long recurrence time scales ($\sim
10^6$ yrs).  That is, the CB disk may need to be reformed after each
nova outburst.  Since the amount of mass required to initiate a
thermonuclear runaway on an intermediate mass white dwarf of $\sim 0.6
M_{\odot}$ is $\sim 10^{-4} M_{\odot}$ (Townsley \& Bildsten 2005), 
about 1\% of this mass should
lie in the CB disk.  Assuming that the disk is destroyed after each
nova event, the rate of mass loss from the donor into the CB disk
would be estimated to be in the range $10^{-13} - 10^{-12}\,
M_{\odot}\, {\rm yr}^{-1}$. On the other hand, if the CB disk is
maintained after the nova outburst the rate of mass loss into the CB
disk can be as low as $10^{-15}\, M_{\odot}\, {\rm yr}^{-1}$.

To conclude, our numerical results suggest that an angular 
momentum loss prescription with properties similar to a CB disk
provides additional ingredients to the angular momentum loss
description, leading to closer agreement with the observed period
distribution below the period gap than has been previously found.
This results from the fact that the angular momentum loss rate is not
only sufficiently enhanced above that due to gravitational radiation
alone, but its rate depends on the properties of the initial binary
when mass transfer is initiated during the CV phase. This leads to the
result that the evolutionary tracks followed by CVs below the period
gap do not converge to a common track. Although other angular 
momentum loss processes with similar characteristics to CB disks are 
possible in principle, their theoretical development has not been 
sufficiently advanced to carry out calculations for meaningful comparisons 
to the observed period distribution.

In summary, the main ingredients which allowed for a good match to the
cumulative distribution function of observed CVs are the inclusion of
an additional angular momentum loss mechanism which does not lead 
to common evolutionary tracks, which in this study is accomplished with 
the effect of a CB disk, a weighting factor to account for observational
selection, and a flow of systems from above the period gap. In the
future, we plan to investigate the orbital period distribution above
the period gap.  Such studies are important for obtaining estimates of
the ratio of systems from above to below the period gap, thus,
providing a measure of the flow rate from long orbital periods to
short periods.  Such constraints can be important for identifying the
cause for the period gap.  

\section{Acknowledgments} 

We are grateful to the anonymous referee whose comments and suggestions led to an improvement of the paper. This research was supported in part by the
National Science Foundation under grants AST 0200876, a David and
Lucille Packard Foundation Fellowship in Science and Engineering
grant, NASA ATP grant NAG5-13236. BW and UK acknowledge the support of
the British Particle Physics and Astronomy Research Council (PPARC).

\clearpage

\clearpage

\clearpage

\clearpage

\clearpage

\clearpage

\clearpage

\clearpage

\clearpage

\clearpage

\clearpage

\begin{deluxetable}{lc}
\tablecolumns{2}
\tablecaption{Population synthesis model parameters. 
   \label{models} }
\tablehead{ 
   \colhead{model} &  
   \colhead{$\alpha_{\rm CE}\, \lambda_{\rm CE}$} 
   }
\startdata
A     & 0.5 \\
CE1   & 0.1 \\
CE8   & 2.5 \\
DCE1  & 0.5 for case B RLO \\
      & 2.5 for case C RLO \\
DCE5  & 2.5 when $M_{\rm e} < 2\,M_2$ \\
      & 0.5 when $M_{\rm e} \ge 2\,M_2$ \\
\enddata
\end{deluxetable}

\clearpage

\begin{deluxetable}{lccccccccc}
\tablecolumns{5}
\tabletypesize{\scriptsize}
\tablecaption{Present-day birth rates (in units of numbers of systems
  per Gyr) of CVs forming at orbital periods below 2.75h, and their
  decomposition according to the type of WD in the system. The
  fractions of systems forming with different types of WDs is indicated
  between parentheses.
\label{br}}
\tablehead{
   \colhead{model} & \colhead{He WD} & \colhead{C/O WD} &
   \colhead{O/Ne/Mg WD} & \colhead{Total} 
   }
\startdata
\multicolumn{5}{l}{\underline{$n(q) \propto q$, $0 < q \le 1$}} \\
\\
A    & $4.8 \times 10^5$ (77.7\%) & $1.4 \times 10^5$ (22.2\%) & $5.7 \times 10^2$ (0.1\%) & $6.1 \times 10^5$ \\
CE1  & $1.7 \times 10^5$ (54.9\%) & $1.4 \times 10^5$ (45.1\%) & $0.0 \times 10^0$ (0.0\%) & $3.2 \times 10^5$ \\
CE8  & $3.4 \times 10^5$ (86.5\%) & $5.2 \times 10^4$ (13.5\%) & $2.6 \times 10^2$ (0.1\%) & $3.9 \times 10^5$ \\
DCE1 & $4.8 \times 10^5$ (90.4\%) & $5.1 \times 10^4$  (9.6\%) & $2.6 \times 10^2$ ($<$0.1\%) & $5.3 \times 10^5$ \\
DCE5 & $4.8 \times 10^5$ (77.5\%) & $1.4 \times 10^5$ (22.4\%) & $5.8 \times 10^2$ (0.1\%) & $6.2 \times 10^5$ \\
\\
\multicolumn{5}{l}{\underline{$n(q) = 1$, $0 < q \le 1$}} \\
\\
A    & $1.6 \times 10^6$ (65.8\%) & $8.1 \times 10^5$ (33.9\%) & $9.3 \times 10^3$ (0.4\%) & $2.4 \times 10^6$ \\
CE1  & $4.7 \times 10^5$ (46.6\%) & $5.4 \times 10^5$ (53.4\%) & $0.0 \times 10^0$ (0.0\%) & $1.0 \times 10^6$ \\
CE8  & $1.3 \times 10^6$ (76.8\%) & $4.0 \times 10^5$ (23.0\%) & $4.9 \times 10^3$ (0.3\%) & $1.7 \times 10^6$ \\
DCE1 & $1.6 \times 10^6$ (80.0\%) & $3.9 \times 10^5$ (19.7\%) & $4.9 \times 10^3$ (0.2\%) & $2.0 \times 10^6$ \\
DCE5 & $1.6 \times 10^6$ (65.3\%) & $8.3 \times 10^5$ (34.3\%) & $9.3 \times 10^3$ (0.4\%) & $2.4 \times 10^6$ \\
\\
\multicolumn{5}{l}{\underline{$n(q) \propto q^{-0.99}$, $0 < q \le 1$}} \\
\\
A    & $1.2 \times 10^5$ (50.2\%) & $1.1 \times 10^5$ (48.4\%) & $3.2 \times 10^3$ (1.4\%) & $2.3 \times 10^5$ \\
CE1  & $2.8 \times 10^4$ (38.6\%) & $4.4 \times 10^4$ (61.4\%) & $0.0 \times 10^0$ (0.0\%) & $7.2 \times 10^4$ \\
CE8  & $1.2 \times 10^5$ (62.4\%) & $6.8 \times 10^4$ (36.6\%) & $2.0 \times 10^3$ (1.1\%) & $1.8 \times 10^5$ \\
DCE1 & $1.2 \times 10^5$ (63.0\%) & $6.6 \times 10^4$ (35.9\%) & $2.0 \times 10^3$ (1.1\%) & $1.8 \times 10^5$ \\
DCE5 & $1.2 \times 10^5$ (49.6\%) & $1.1 \times 10^5$ (49.0\%) & $3.2 \times 10^3$ (1.4\%) & $2.3 \times 10^5$ \\
\\
\multicolumn{5}{l}{\underline{$M_2$ from same IMF as $M_1$}} \\
\\
A    & $5.8 \times 10^6$ (49.9\%) & $5.7 \times 10^6$ (48.7\%) & $1.6 \times 10^5$ (1.3\%) & $1.2 \times 10^7$ \\
CE1  & $1.4 \times 10^6$ (38.1\%) & $2.2 \times 10^6$ (61.9\%) & $0.0 \times 10^0$ (0.0\%) & $3.6 \times 10^6$ \\
CE8  & $6.0 \times 10^6$ (62.1\%) & $3.5 \times 10^6$ (36.8\%) & $1.0 \times 10^5$ (1.1\%) & $9.6 \times 10^6$ \\
DCE1 & $5.8 \times 10^6$ (62.0\%) & $3.5 \times 10^6$ (36.9\%) & $1.0 \times 10^5$ (1.1\%) & $9.4 \times 10^6$ \\
DCE5 & $5.8 \times 10^6$ (49.4\%) & $5.8 \times 10^6$ (49.3\%) & $1.6 \times 10^5$ (1.3\%) & $1.2 \times 10^7$ \\
\enddata 
\end{deluxetable}

\clearpage

\begin{deluxetable}{lccccccccc}
\tablecolumns{10}
\setlength{\tabcolsep}{0.02in}
\tabletypesize{\scriptsize}
\tablecaption{Total number of CVs with orbital periods shorter than
  2.75\,hrs currently populating the Galactic disk. The numbers only
  reflect systems that were formed below 2.75\,hrs and thus do not
  account for systems evolving through the period gap from periods
  longer than 2.75\,hrs. The relative contributions of systems with
  He, C/O, and O/Ne/Mg WDs to the population are indicated between
  parentheses. Approximate space densities can be obtained by dividing
  the absolute numbers of systems by $5 \times 10^{11}\,{\rm pc^3}$
  (see Willems \& Kolb 2004). 
\label{num}}
\tablehead{
   \colhead{} & \multicolumn{4}{c}{Without CB disk} &  & 
   \multicolumn{4}{c}{With CB disk} \\
   \cline{2-5} \cline{7-10}
   \colhead{model} & \colhead{He WD} & \colhead{C/O WD} &
   \colhead{O/Ne/Mg WD} & \colhead{Total} &  & \colhead{He WD} &
   \colhead{C/O WD} & \colhead{O/Ne/Mg WD} & \colhead{Total}
   }
\startdata
\multicolumn{10}{l}{\underline{$n(q) \propto q$, $0 < q \le 1$}} \\
\\
A    & $7.1 \times 10^6 \,\,\,(69.7\%)$ & $3.1 \times 10^6 \,\,\,(30.2\%)$ & $1.2 \times 10^4 \,\,\,(0.1\%)$ & $1.0 \times 10^7$ &  & $4.3 \times 10^6 \,\,\,(73.1\%)$ & $1.6 \times 10^6 \,\,\,(26.8\%)$ & $5.0 \times 10^3 \,\,\,(0.1\%)$ & $5.9 \times 10^6$ \\
CE1  & $2.2 \times 10^6 \,\,\,(45.7\%)$ & $2.7 \times 10^6 \,\,\,(54.3\%)$ & $0.0 \times 10^0 \,\,\,(0.0\%)$ & $4.9 \times 10^6$ &  & $1.7 \times 10^6 \,\,\,(51.7\%)$ & $1.6 \times 10^6 \,\,\,(48.3\%)$ & $0.0 \times 10^0 \,\,\,(0.0\%)$ & $3.4 \times 10^6$ \\
CE8  & $4.1 \times 10^6 \,\,\,(80.9\%)$ & $9.5 \times 10^5 \,\,\,(19.0\%)$ & $5.0 \times 10^3 \,\,\,(0.1\%)$ & $5.0 \times 10^6$ &  & $2.1 \times 10^6 \,\,\,(79.5\%)$ & $5.4 \times 10^5 \,\,\,(20.4\%)$ & $2.2 \times 10^3 \,\,\,(0.1\%)$ & $2.6 \times 10^6$ \\
DCE1 & $7.1 \times 10^6 \,\,\,(89.3\%)$ & $8.4 \times 10^5 \,\,\,(10.6\%)$ & $5.0 \times 10^3 \,\,\,(0.1\%)$ & $7.9 \times 10^6$ &  & $4.3 \times 10^6 \,\,\,(89.6\%)$ & $5.0 \times 10^5 \,\,\,(10.4\%)$ & $2.2 \times 10^3 \,\,\,(<0.1\%)$ & $4.8 \times 10^6$ \\
DCE5 & $7.1 \times 10^6 \,\,\,(69.3\%)$ & $3.1 \times 10^6 \,\,\,(30.6\%)$ & $1.3 \times 10^4 \,\,\,(0.1\%)$ & $1.0 \times 10^7$ &  & $4.3 \times 10^6 \,\,\,(72.7\%)$ & $1.6 \times 10^6 \,\,\,(27.2\%)$ & $5.2 \times 10^3 \,\,\,(0.1\%)$ & $5.9 \times 10^6$ \\
\\
\multicolumn{10}{l}{\underline{$n(q) = 1$, $0 < q \le 1$}} \\
\\
A    & $2.3 \times 10^7 \,\,\,(58.8\%)$ & $1.6 \times 10^7 \,\,\,(40.7\%)$ & $1.9 \times 10^5 \,\,\,(0.5\%)$ & $3.9 \times 10^7$ &  & $1.3 \times 10^7 \,\,\,(62.7\%)$ & $7.8 \times 10^6 \,\,\,(37.0\%)$ & $7.3 \times 10^4 \,\,\,(0.3\%)$ & $2.1 \times 10^7$ \\
CE1  & $5.4 \times 10^6 \,\,\,(37.3\%)$ & $9.0 \times 10^6 \,\,\,(62.7\%)$ & $0.0 \times 10^0 \,\,\,(0.0\%)$ & $1.4 \times 10^7$ &  & $4.1 \times 10^6 \,\,\,(43.6\%)$ & $5.3 \times 10^6 \,\,\,(56.4\%)$ & $0.0 \times 10^0 \,\,\,(0.0\%)$ & $9.5 \times 10^6$ \\
CE8  & $1.7 \times 10^7 \,\,\,(72.0\%)$ & $6.5 \times 10^6 \,\,\,(27.7\%)$ & $8.2 \times 10^4 \,\,\,(0.3\%)$ & $2.3 \times 10^7$ &  & $8.5 \times 10^6 \,\,\,(71.4\%)$ & $3.4 \times 10^6 \,\,\,(28.3\%)$ & $3.5 \times 10^4 \,\,\,(0.3\%)$ & $1.2 \times 10^7$ \\
DCE1 & $2.3 \times 10^7 \,\,\,(79.4\%)$ & $5.8 \times 10^6 \,\,\,(20.3\%)$ & $8.1 \times 10^4 \,\,\,(0.3\%)$ & $2.9 \times 10^7$ &  & $1.3 \times 10^7 \,\,\,(80.6\%)$ & $3.1 \times 10^6 \,\,\,(19.1\%)$ & $3.5 \times 10^4 \,\,\,(0.2\%)$ & $1.6 \times 10^7$ \\
DCE5 & $2.3 \times 10^7 \,\,\,(58.1\%)$ & $1.6 \times 10^7 \,\,\,(41.4\%)$ & $1.9 \times 10^5 \,\,\,(0.5\%)$ & $3.9 \times 10^7$ &  & $1.3 \times 10^7 \,\,\,(62.1\%)$ & $8.0 \times 10^6 \,\,\,(37.5\%)$ & $7.5 \times 10^4 \,\,\,(0.4\%)$ & $2.1 \times 10^7$ \\
\\
\multicolumn{10}{l}{\underline{$n(q) \propto q^{-0.99}$, $0 < q \le 1$}} \\
\\
A    & $1.6 \times 10^6 \,\,\,(44.7\%)$ & $2.0 \times 10^6 \,\,\,(53.6\%)$ & $6.2 \times 10^4 \,\,\,(1.7\%)$ & $3.6 \times 10^6$ &  & $8.9 \times 10^5 \,\,\,(49.3\%)$ & $9.0 \times 10^5 \,\,\,(49.4\%)$ & $2.2 \times 10^4 \,\,\,(1.2\%)$ & $1.8 \times 10^6$ \\
CE1  & $2.8 \times 10^5 \,\,\,(29.3\%)$ & $6.7 \times 10^5 \,\,\,(70.7\%)$ & $0.0 \times 10^0 \,\,\,(0.0\%)$ & $9.4 \times 10^5$ &  & $2.1 \times 10^5 \,\,\,(35.9\%)$ & $3.8 \times 10^5 \,\,\,(64.1\%)$ & $0.0 \times 10^0 \,\,\,(0.0\%)$ & $5.9 \times 10^5$ \\
CE8  & $1.5 \times 10^6 \,\,\,(58.8\%)$ & $1.0 \times 10^6 \,\,\,(40.0\%)$ & $3.0 \times 10^4 \,\,\,(1.2\%)$ & $2.5 \times 10^6$ &  & $7.3 \times 10^5 \,\,\,(59.5\%)$ & $4.9 \times 10^5 \,\,\,(39.5\%)$ & $1.2 \times 10^4 \,\,\,(1.0\%)$ & $1.2 \times 10^6$ \\
DCE1 & $1.6 \times 10^6 \,\,\,(62.8\%)$ & $9.4 \times 10^5 \,\,\,(36.0\%)$ & $3.0 \times 10^4 \,\,\,(1.2\%)$ & $2.6 \times 10^6$ &  & $8.9 \times 10^5 \,\,\,(65.7\%)$ & $4.6 \times 10^5 \,\,\,(33.4\%)$ & $1.2 \times 10^4 \,\,\,(0.9\%)$ & $1.4 \times 10^6$ \\
DCE5 & $1.6 \times 10^6 \,\,\,(43.9\%)$ & $2.0 \times 10^6 \,\,\,(54.4\%)$ & $6.3 \times 10^4 \,\,\,(1.7\%)$ & $3.7 \times 10^6$ &  & $8.9 \times 10^5 \,\,\,(48.6\%)$ & $9.2 \times 10^5 \,\,\,(50.2\%)$ & $2.3 \times 10^4 \,\,\,(1.2\%)$ & $1.8 \times 10^6$ \\
\\
\multicolumn{10}{l}{\underline{$M_2$ from same IMF as $M_1$}} \\
\\
A    & $8.0 \times 10^7 \,\,\,(44.6\%)$ & $9.7 \times 10^7 \,\,\,(53.7\%)$ & $3.1 \times 10^6 \,\,\,(1.7\%)$ & $1.8 \times 10^8$ &  & $4.4 \times 10^7 \,\,\,(49.5\%)$ & $4.3 \times 10^7 \,\,\,(49.2\%)$ & $1.1 \times 10^6 \,\,\,(1.2\%)$ & $8.8 \times 10^7$ \\
CE1  & $1.3 \times 10^7 \,\,\,(28.8\%)$ & $3.2 \times 10^7 \,\,\,(71.2\%)$ & $0.0 \times 10^0 \,\,\,(0.0\%)$ & $4.5 \times 10^7$ &  & $9.9 \times 10^6 \,\,\,(35.5\%)$ & $1.8 \times 10^7 \,\,\,(64.5\%)$ & $0.0 \times 10^0 \,\,\,(0.0\%)$ & $2.8 \times 10^7$ \\
CE8  & $7.7 \times 10^7 \,\,\,(59.1\%)$ & $5.2 \times 10^7 \,\,\,(39.7\%)$ & $1.5 \times 10^6 \,\,\,(1.2\%)$ & $1.3 \times 10^8$ &  & $3.7 \times 10^7 \,\,\,(60.2\%)$ & $2.4 \times 10^7 \,\,\,(38.8\%)$ & $6.2 \times 10^5 \,\,\,(1.0\%)$ & $6.2 \times 10^7$ \\
DCE1 & $8.0 \times 10^7 \,\,\,(61.8\%)$ & $4.8 \times 10^7 \,\,\,(37.0\%)$ & $1.5 \times 10^6 \,\,\,(1.2\%)$ & $1.3 \times 10^8$ &  & $4.4 \times 10^7 \,\,\,(65.0\%)$ & $2.3 \times 10^7 \,\,\,(34.0\%)$ & $6.2 \times 10^5 \,\,\,(0.9\%)$ & $6.7 \times 10^7$ \\
DCE5 & $8.0 \times 10^7 \,\,\,(43.8\%)$ & $1.0 \times 10^8 \,\,\,(54.5\%)$ & $3.1 \times 10^6 \,\,\,(1.7\%)$ & $1.8 \times 10^8$ &  & $4.4 \times 10^7 \,\,\,(48.8\%)$ & $4.5 \times 10^7 \,\,\,(50.0\%)$ & $1.1 \times 10^6 \,\,\,(1.2\%)$ & $9.0 \times 10^7$ \\
\enddata 
\end{deluxetable}

\clearpage

\begin{deluxetable}{lccccccccc}
\tablecolumns{10}
\tabletypesize{\scriptsize}
\tablecaption{Intrinsic fractions of CVs forming below
  2.75h that are still evolving towards the period minimum (pre-bounce
  systems) and that are evolving away from the period minimum
  (post-bounce systems), and their decomposition according to the type of WD in the system. The
  fractions are expressed in per cent with the first number in each
  column corresponding to pre-bounce systems and the second to
  post-bounce systems. 
\label{prepost}}
\tablehead{
   \colhead{} & \multicolumn{4}{c}{Without CB disk} &  & 
   \multicolumn{4}{c}{With CB disk} \\
   \cline{2-5} \cline{7-10}
   \colhead{model} & \colhead{He WD} & \colhead{C/O WD} &
   \colhead{O/Ne/Mg WD} & \colhead{Total} &  & \colhead{He WD} &
   \colhead{C/O WD} & \colhead{O/Ne/Mg WD} & \colhead{Total}
   }
\startdata
\multicolumn{10}{l}{\underline{$n(q) \propto q$, $0 < q \le 1$}} \\
\\
A    & 61.1/9.0  & 19.3/10.6 & $<$0.1/0.1    & 80.4/19.7 &  & 59.5/13.5 & 21.1/5.8  & 0.1/$<$0.1    & 80.7/19.3 \\
CE1  & 44.8/0.7  & 40.9/13.5 &    0.0/0.0    & 85.7/14.2 &  & 45.1/6.5  & 38.6/9.8  & 0.0/0.0       & 83.7/16.3 \\
CE8  & 62.2/19.6 & 11.7/6.4  & $<$0.1/0.1    & 73.9/26.1 &  & 59.7/19.8 & 15.6/4.8  & 0.1/$<$0.1    & 75.4/24.6 \\
DCE1 & 78.4/11.6 &  6.5/3.6  & $<$0.1/$<$0.1 & 84.9/15.2 &  & 73.1/16.6 &  7.9/2.5  & $<$0.1/$<$0.1 & 81.0/19.1 \\
DCE5 & 60.7/8.9  & 19.6/10.7 & $<$0.1/0.1    & 80.3/19.7 &  & 59.3/13.5 & 21.3/5.9  & 0.1/$<$0.1    & 80.7/19.4 \\
\\
\multicolumn{10}{l}{\underline{$n(q) = 1$, $0 < q \le 1$}} \\
\\
A    & 47.8/11.3 & 22.7/17.8 & 0.2/0.3 & 70.7/29.4 &  & 48.9/13.9 & 27.3/9.7  & 0.2/0.1 & 76.4/23.7 \\
CE1  & 36.1/1.1  & 43.3/19.5 & 0.0/0.0 & 79.4/20.6 &  & 37.2/6.5  & 43.2/13.1 & 0.0/0.0 & 80.4/19.6 \\
CE8  & 50.9/21.8 & 14.7/12.4 & 0.1/0.2 & 65.7/34.4 &  & 51.1/20.2 & 20.2/8.2  & 0.2/0.1 & 71.5/28.5 \\
DCE1 & 64.7/15.2 & 10.5/9.2  & 0.1/0.2 & 75.3/24.6 &  & 62.8/17.8 & 13.5/5.7  & 0.1/0.1 & 76.4/23.6 \\
DCE5 & 47.3/11.2 & 23.0/18.1 & 0.2/0.3 & 70.5/29.6 &  & 48.4/13.7 & 27.7/9.8  & 0.2/0.1 & 76.3/23.6 \\
\\
\multicolumn{10}{l}{\underline{$n(q) \propto q^{-0.99}$, $0 < q \le 1$}} \\
\\
A    & 32.9/12.2 & 24.5/29.0 & 0.4/1.1 & 57.8/42.3 &  & 36.1/13.2 & 33.3/16.2 & 0.7/0.4 & 70.1/29.8 \\
CE1  & 27.7/1.5  & 43.7/27.1 & 0.0/0.0 & 71.4/28.6 &  & 29.4/6.3  & 46.6/17.6 & 0.0/0.0 & 76.0/23.9 \\
CE8  & 37.6/21.8 & 17.7/22.0 & 0.2/0.7 & 55.5/44.5 &  & 40.2/19.3 & 25.5/14.1 & 0.5/0.5 & 66.2/33.9 \\
DCE1 & 46.2/17.2 & 15.5/20.3 & 0.2/0.7 & 61.9/38.2 &  & 48.0/17.7 & 21.2/12.3 & 0.5/0.4 & 69.7/30.4 \\
DCE5 & 32.3/11.9 & 24.9/29.5 & 0.4/1.1 & 57.6/42.5 &  & 35.5/13.1 & 33.9/16.4 & 0.7/0.4 & 70.1/29.9 \\
\\
\multicolumn{10}{l}{\underline{$M_2$ from same IMF as $M_1$}} \\
\\
A    & 31.7/13.3 & 23.0/30.5 & 0.4/1.1 & 55.1/44.9 &  & 35.4/14.1 & 32.1/17.3 & 0.7/0.5 & 68.2/31.9 \\
CE1  & 26.9/1.7  & 42.3/29.1 & 0.0/0.0 & 69.2/30.8 &  & 28.7/6.7  & 45.8/18.9 & 0.0/0.0 & 74.5/25.6 \\
CE8  & 36.4/23.2 & 16.3/22.9 & 0.2/0.8 & 52.9/46.9 &  & 39.8/20.3 & 24.2/14.7 & 0.5/0.5 & 64.5/35.5 \\
DCE1 & 43.9/18.5 & 14.8/21.9 & 0.2/0.8 & 58.9/41.2 &  & 46.5/18.5 & 20.8/13.3 & 0.5/0.4 & 67.8/32.2 \\
DCE5 & 31.1/13.0 & 23.4/31.0 & 0.4/1.1 & 54.9/45.1 &  & 34.9/13.9 & 32.6/17.4 & 0.7/0.5 & 68.2/31.8 \\
\enddata 
\end{deluxetable}

\end{document}